\documentclass{article}
\usepackage{graphicx} 
\usepackage{amsmath}
\usepackage{amssymb}
\usepackage[margin=1in]{geometry}
\usepackage{cite}
\usepackage{hyperref}
\usepackage{cleveref}

\begin{document}

\begin{center}
    {\Large
    \textbf{A Direct Measurement of Hard Two-Photon Exchange with Electrons and Positrons at CLAS12}
    }\\
    \vspace{5pt}
    {\large
    A Proposal to Jefferson Lab PAC 51
    }
    \vspace{15pt}

    A.~Schmidt (contact person), W.~J. Briscoe, O.~Cortes, L.~Earnest, G.~N.~Grauvogel, S. Ratliff, E.~M.~Seroka, P.~Sharp, I.~I.~Strakovsky\\
    \emph{The George Washington University}
    \vspace{11pt}

    G.~Niculescu\\
    \emph{James Madison University}
    \vspace{11pt}

    S.~Diehl\\
    \emph{Justus Liebig University Giessen and University of Connecticut}
    \vspace{11pt}
    
    P.~G.~Blunden\\
    \emph{University of Manitoba}
    \vspace{11pt}

    E.~Cline\footnote{Also with Stony Brook University} (spokesperson), I.~Korover (spokesperson), T.~Kutz (spokesperson)\\
    \emph{Massachusetts Institute of Technology}
    \vspace{11pt}

    S.~N.~Santiesteban (spokesperson)\\
    \emph{University of New Hampshire}
    \vspace{11pt}

    C.~Fogler, L.~B.~Weinstein\\
    \emph{Old Dominion University}
    \vspace{11pt}

    D.~Marchand, S.~Niccolai, E.~Voutier\\
    \emph{Université Paris-Saclay, CNRS/IN2P3, IJCLab}
    \vspace{11pt}

    A.~D'Angelo\\
    \emph{INFN Roma Tor Vergata and University of Rome Tor Vergata}
    \vspace{11pt}

    J.~C.~Bernauer\footnote{Also with RIKEN BNL Research Center} (spokesperson), R.~Singh\\
    \emph{Stony Brook University}
    \vspace{11pt}
    
    V.~Burkert (spokesperson), F.~Hauenstein, D.~W.~Higinbotham, D.~Nguyen, E.~Pasyuk, H.~Szumila-Vance, X.~Wei\\
    \emph{Thomas Jefferson National Accelerator Facility}
    \vspace{11pt}

    D.~Keller\\
    \emph{University of Virginia}
    \vspace{20pt}

    A CLAS Collaboration and Jefferson Lab Positron Working Group Proposal\\

    \vspace{35pt}
    May 22, 2023
    
\end{center}
\clearpage

\begin{center}
    \textbf{Executive Summary}
\end{center}

  One of the most surprising discoveries made at Jefferson Lab has been the discrepancy
  in the determinations of the proton's form factor ratio $\mu_p G_E^p/G_M^p$ between
  unpolarized cross section measurements and the polarization transfer technique. 
  Over two decades later, the discrepancy not only persists but has been confirmed
  at higher momentum transfers now accessible in the 12-GeV era. 
  The leading hypothesis for the cause of this discrepancy, a non-negligible contribution
  from hard two-photon exchange, has neither been conclusively proven or disproven.
  This state of uncertainty not only clouds our knowledge of one-dimensional nucleon structure
  but also poses a major concern for our field's efforts to map out the three-dimensional nuclear
  structure. A better understanding of multi-photon exchange over a wide phase space is needed.

  We propose making comprehensive measurements of two-photon exchange over a wide range
  in momentum transfer and scattering angle using the CLAS12 detector. Specifically, we
  will measure the ratio of positron-proton to electron-proton elastic scattering
  cross sections, using the proposed positron beam upgrade for CEBAF. The experiment will
  use 2.2, 4.4, and 6.6 GeV lepton beams incident on the standard CLAS12 unpolarized hydrogen
  target. Data will be collected by the CLAS12 detector in its standard configuration, except for
  a modified trigger to allow the recording of events with beam leptons scattered into
  the CLAS12 central detector.
  The sign of the beam charge, as well as the polarity of the CLAS12 solenoid and toroid, will be reversed
  several times in order to suppress systematics associated with local detector efficiency
  and time-dependent detector performance.
  
  The proposed high-precision determination of two-photon effects will be a definitive test
  of whether or not hard two-photon exchange is the root cause of the proton form factor discrepancy.
  The measurements will also provide strong constraints to help guide theoretical efforts, particularly
  in kinematics that pose challenges to current theoretical approaches.
  We request 55 PAC days.

\clearpage

\tableofcontents

\section{Introduction}

\begin{figure}[htb]
    \centering
    \includegraphics[width=\textwidth]{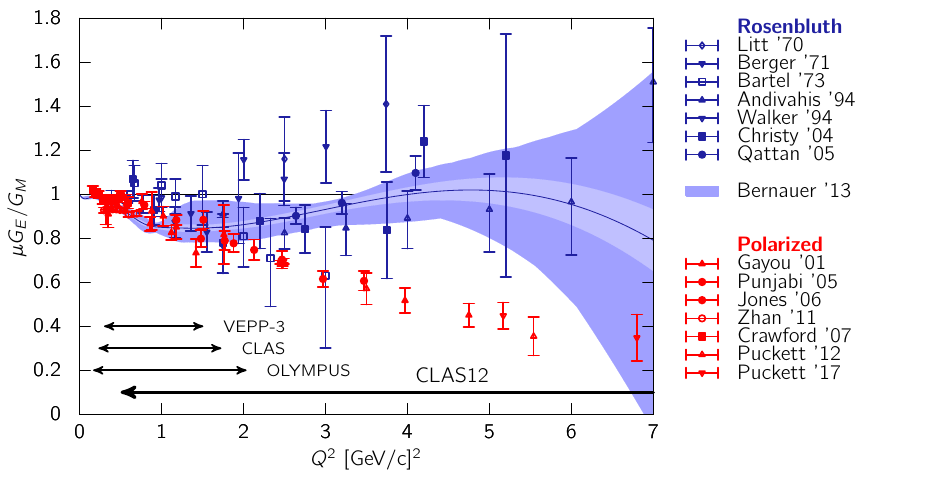}
    \caption{A subset of world data on the proton's form factor ratio $\mu_p G_E^p / G_M^p$ show the discrepancy between Rosenbluth and polarized measurements. 
    Rosenbluth separations of unpolarized cross sections (blue points) are taken from Refs.~\cite{Litt:1969my,Berger:1971kr,Bartel:1973rf,Andivahis:1994rq,E94110:2004lsx,Qattan:2004ht}. Polarized measurements (red points) are taken from Refs.~\cite{Gayou:2001qt,Punjabi:2005wq,ResonanceSpinStructure:2006oim,Zhan:2011ji,Crawford:2006rz,Puckett:2011xg,Puckett:2017flj}. 
    A global fit to unpolarized cross sections along with estimates of statistical and total uncertainty by Bernauer et al.~\cite{A1:2013fsc} are shown as blue bands. For comparison, the $Q^2$ coverage of recent two-photon exchange measurements by VEPP-3~\cite{Rachek:2014fam}, CLAS~\cite{CLAS:2016fvy}, and OLYMPUS~\cite{OLYMPUS:2016gso} are shown with arrows. The coverage of this proposal is also shown, labeled `CLAS12.'
    }
    \label{fig:GE_GM}
\end{figure}

The proton's electromagnetic form factors, $G_E^p$ and $G_M^p$ are fundamental properties that encode the spatial distribution of the proton's charge and current. Decades of measurements mapping out the proton's form factors were suddenly upended by new experiments at Jefferson Lab and elsewhere that used polarization observables, rather than unpolarized cross section measurements, to directly access the form factor ratio $\mu_p G_E^p / G_M^p$. Whereas Rosenbluth separations of unpolarized cross section data generally suggest that $\mu_p G_E^p / G_M^p$ is approximately 1, the polarized measurements show an unambiguous downwards slope as a function of squared four-momentum transfer $Q^2$. A subset of world data~\cite{Litt:1969my,Berger:1971kr,Bartel:1973rf,Andivahis:1994rq,E94110:2004lsx,Qattan:2004ht,Gayou:2001qt,Punjabi:2005wq,ResonanceSpinStructure:2006oim,Zhan:2011ji,Crawford:2006rz,Puckett:2011xg,Puckett:2017flj}, showing this discrepancy is found in Fig.~\ref{fig:GE_GM}. As Rosenbluth separations invariably produce large coupled uncertainties in $G_E$  and $G_M$, a more modern technique is to perform a global fit directly to unpolarized cross section measurements. An example, from Ref.~\cite{A1:2013fsc}, is shown as well, along with uncertainty bands. The discrepancy is most prominent between $Q^2$ values of approximately 2--7 GeV$^2/c^2$ where there is a clear gap between the polarized and unpolarized data. At higher $Q^2$, the large uncertainties in the unpolarized data obscure whether the discrepancy persists at even higher $Q^2$, though recent results from Hall A suggest that it does~\cite{Christy:2021snt}.

\begin{figure}[htb]
    \centering
    \includegraphics[width=0.95\textwidth]{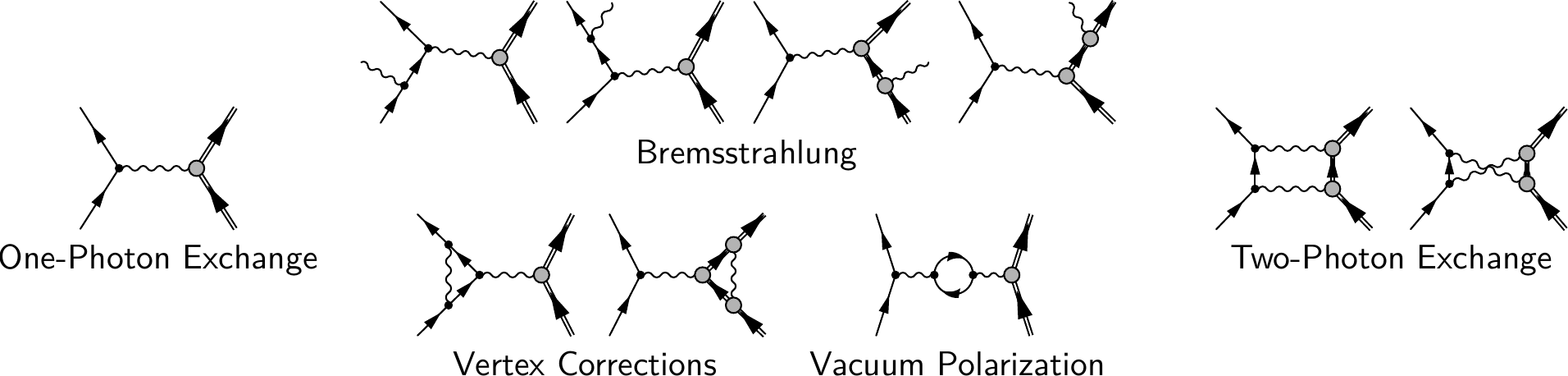}
    \caption{Diagrams treated in standard radiative corrections to elastic electron scattering}
    \label{fig:diagrams}
\end{figure}

A leading hypothesis for the source of the discrepancy quickly emerged: a radiative effect called hard two-photon exchange, which is neglected in standard radiative corrections formulae, may be biasing the two measurement techniques differently~\cite{Guichon:2003qm,Blunden:2003sp}. Radiative correction procedures, such as that of Mo and Tsai~\cite{Mo:1968cg}, or Maximon and Tjon~\cite{Maximon:2000hk}, treat two-photon exchange (TPE) in the so-called ``soft-limit,'' an approximation in which one of the photons carries negligible four-momentum. This approximation makes the TPE diagram tractable without model-dependence. The TPE contribution beyond the soft-limit is referred to as ``hard two-photon exchange,'' or hard TPE. It should be noted that two-photon exchange has to be treated in order to cancel divergences in bremsstrahlung terms. The set of diagrams treated by standard radiative corrections procedures are shown in Fig.~\ref{fig:diagrams}. 

The hypothesis that hard TPE could produce the proton form factor discrepancy led to a renewed interest in attempting to quantify the effect experimentally. One of the most straight-forward experimental signatures is to observe a difference in cross sections in positron scattering relative to electron scattering. The lowest order TPE correction to the elastic scattering cross section is an interference term between one-photon exchange (OPE) and two-photon exchange. This interference term is proportional to $e^3$, where $e$ is the lepton charge, and therefore changes sign between positron and electron scattering. This can be written:
\begin{equation}
\sigma \sim |\mathcal{M}_{1\gamma}|^2 \pm 2\text{Re}\left[ \mathcal{M}_{1\gamma} \mathcal{M}_{2\gamma}\right] + \ldots,
\end{equation}
where $\mathcal{M}_{1\gamma}$ and $\mathcal{M}_{2\gamma}$ are the OPE and TPE matrix elements. The positron-proton to electron-proton cross section ratio, denoted $R_{2\gamma}$, is therefore:
\begin{equation}
    R_{2\gamma} \equiv \frac{\sigma_{e^+p}}{\sigma_{e^-p}} = 1 + 4\frac{\text{Re}\left[ \mathcal{M}_{1\gamma} \mathcal{M}_{2\gamma}\right]}{|\mathcal{M}_{1\gamma}|^2} + \ldots
\end{equation}

This should be interpreted as a sketch of the principle of the measurement. In practice, one must account for all of the other radiative effects entering at the same order, most of which are lepton charge-symmetric but some (e.g., lepton-proton bremsstrahlung interference) are not. 

Three experiments were proposed in response to the TPE hypothesis, and collected data in 2010--2013. The results of those experiments were not conclusive, primarily because the experiments lacked the kinematic reach necessary to probe the region of $Q^2$ where the discrepancy is significant. Gathering data with the necessary precision and kinematic reach to conclusively test if TPE is the cause of the proton form factor discrepancy is the main motivation for this proposal. In the following sections we briefly review recent theoretical, phenomenological, and experimental efforts. For more details we refer to Ref.~\cite{Afanasev:2017gsk}.

For what follows, it is helpful to recall that in elastic scattering at fixed beam energy, there is only one kinematic degree-of-freedom. Therefore, the kinematics of any elastic scattering measurement can be represented by just two numbers, for example beam energy and scattering angle. We will frequently use the combination $Q^2$ and $\epsilon$, where $Q^2$ is the familiar negative squared four-momentum transfer, given by:
\begin{align}
    Q^2 = & \; 2EE'(1-\cos\theta) \\
    \rightarrow  & \; \frac{2E^2 m_p (1-\cos\theta)}{m_p + E(1-\cos\theta)},
\end{align}
where $E$ is the beam energy, $E'$ is the outgoing lepton energy, $m_p$ is the proton mass, and $\theta$ is the lepton scattering angle; and $\epsilon$ is the 
virtual photon longitudinal polarization parameter, given in elastic kinematics by:
\begin{equation}
    \epsilon \rightarrow \left[ 1+ 2(1+\tau)\tan^2\frac{\theta}{2}\right]^{-1},
\end{equation}
where $\tau \equiv \frac{Q^2}{4m_p^2}$.
Increasing the lepton scattering angle, $\theta$, increases $Q^2$ and decreases $\epsilon$.

\subsection{Two-Photon Exchange Theory}

One of the challenges preventing incorporation of hard TPE into standard radiative corrections prescriptions is difficulty in calculating the TPE diagram without adding significant model dependence. 
The TPE diagram has an off-shell hadronic propagator, which requires some QCD input in order to be evaluated. 
In general, there are two classes of approaches: hadronic methods and partonic methods, which are described in the following sections. There are also a handful of alternative theoretical approaches which fit in neither category, some of which suggest that TPE cannot be the cause of the form factor discrepancy (e.g.~\cite{Kuraev:2007dn}).

Efforts to advance our understanding of TPE and of radiative corrections more generally is an active area of research, with applications beyond merely the discrepancy in the proton form factors. TPE is a significant correction in scattering experiments at low $Q^2$ to determine the proton radius. The TPE diagram shares many of the same challenges with the $\gamma Z$ box diagram that is a major radiative correction in parity-violating electron scattering, and with the $\gamma W$ box diagram in nuclear $\beta$-decay. An Ad-Hoc Workshop on Radiative Corrections, held (remotely) in 2020 by CFNS at Stony Brook produced a white paper~\cite{Afanasev:2020hwg} on current challenges in radiative corrections. This was followed by a workshop at the ECT in Trento in 2022 with a second paper forthcoming \cite{radcorwp}.  

\subsubsection{Hadronic Methods}

In hadronic methods, the off-shell propagator is expanded as a sum of on-shell intermediate hadronic states. This series is, in principle, infinite, containing all baryons with quantum numbers accessible from photon-nucleon coupling. In practice, the higher mass resonances contribute less and less with mass, and the expansion can be truncated to get an approximate result. Additional model dependence comes from the assumptions about the transition form factors for each intermediate state. Early works used the direct evaluation of loop integrals~\cite{Blunden:2003sp,Blunden:2005ew}, but the flexibility of hadronic calculations have greatly improved by using dispersive methods~\cite{Gorchtein:2006mq,Borisyuk:2008es,Borisyuk:2015xma,Blunden:2017nby,Ahmed:2020uso}. The most advanced dispersive hadronic calculation incorporates contributions from the nucleon as well as all $N*$ and $\Delta$ resonances with masses up to 1.8~GeV~\cite{Ahmed:2020uso}. 

Hadronic approaches should perform better at low momentum transfer, and calculations are typically not performed above $Q^2 > 3$~GeV$^2/c^2$. In that limit, however, the results of hadronic calculations suggest that hard TPE would drive the apparent $\mu_p G_E/G_M$ extracted from unpolarized cross sections up and away from the results of polarization transfer~\cite{Blunden:2003sp,Arrington:2007ux}. Nevertheless, hadronic calculations do show some variance based on different model assumptions used and which intermediate states are included. 

\subsubsection{Partonic Methods}

Partonic methods model the hard interaction of the two exchanged photons with individual quarks, but then embed those quarks inside the proton using a model of the proton's partonic structure---either its distribution amplitudes (DAs) or generalized parton distributions (GPDs). This approach is more accurate at high momentum transfers where the factorization between the hard scattering and the soft structure are on firmer ground, ideally above $Q^2>5$~GeV$^2/c^2$. The accuracy is only as good, however, as our understanding of the proton's multi-dimensional structure. Examples of partonic estimates include Refs.~\cite{Chen:2004tw,Afanasev:2005mp,Borisyuk:2008db,Kivel:2009eg,Kivel:2012vs}.

\subsection{Phenomenology}

There is a class of phenomenological two-photon exchange estimates that use experimental data to estimate the magnitude of the two-photon exchange amplitude, making the assumption that hard TPE is the sole cause of the proton form factor discrepancy~\cite{Chen:2007ac,Borisyuk:2010ep,Guttmann:2010au,A1:2013fsc,Schmidt:2019vpr}. This amounts to determining how much of a hard TPE effect would be necessary to fully resolve the discrepancy. If the measured TPE differed substantially from these phenomenological predictions, this would be evidence that factors other than hard TPE are responsible for the discrepancy. However, even modestly different assumptions lead to a wide range of phenomenological estimates, which implies that even the magnitude of the discrepancy is not well-constrained.

\begin{figure}
    \centering
    \includegraphics{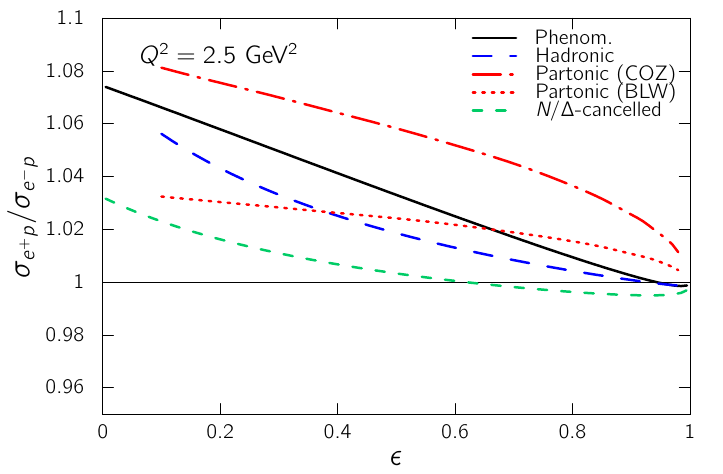}
    \caption{Predictions for $R_{2\gamma}$, the ratio of positron-proton to electron-proton elastic cross sections, shown
    as a function of $\epsilon$, for a fixed $Q^2=2.50$~GeV$^2/c^2$. The phenomenological prediction comes from Bernauer et al.\ (Mainz A1)~\cite{A1:2013fsc}. The hadronic prediction is the $N+\Delta$ calculation of Blunden et al.~(2017) \cite{Blunden:2017nby}. 
    The partonic calculations come from Kivel and Vanderhaeghen (2019)~\cite{Kivel:2009eg}. The calculation labelled 
    $N/\Delta$-cancelled is by Kuraev et al.~(2008) \cite{Kuraev:2007dn}, whose authors conclude that TPE \emph{cannot} be the 
    only cause of the proton form factor discrepancy. The predictions assume the Maximon and Tjon~\cite{Maximon:2000hk} definition
    of soft TPE.}
    \label{fig:tpe_pred_QSq250}
\end{figure}

\begin{figure}
    \centering
    \includegraphics{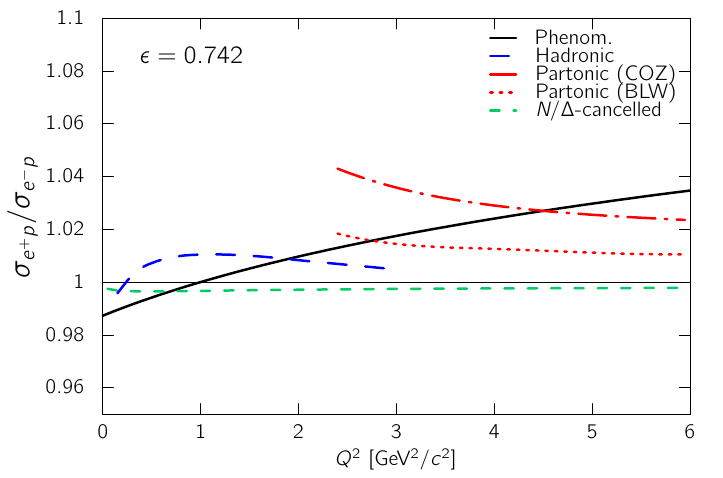}
    \caption{Predictions for $R_{2\gamma}$, the ratio of positron-proton to electron-proton elastic cross sections, shown
    as a function of $Q^2$, for a fixed $\epsilon=0.742$. The calculations are the same as in Fig.~\ref{fig:tpe_pred_QSq250}.}
    \label{fig:tpe_pred_eps742}
\end{figure}

Figures~\ref{fig:tpe_pred_QSq250} and \ref{fig:tpe_pred_eps742} compare several examples of theoretical predictions 
\cite{Blunden:2017nby,Kivel:2009eg,Kuraev:2007dn} (colored curves) against a phenomenological estimate~\cite{A1:2013fsc}
(black curve) by the Mainz A1 collaboration, which assumes that the entire proton form factor discrepancy can be 
attributed to hard TPE, among other assumptions. This curve is a useful reference; theoretical predictions that are similar
generally indicate that TPE could resolve the form factor discrepancy, while theoretical predictions that deviate suggest
that other factors must be involved. In general predictions show that $R_{2\gamma}$ increases as $\epsilon$ decreases.
The hadronic calculation of Blunden et al.\ (2017) \cite{Blunden:2017nby} is below the phenomenological curve at $Q^2=2.5$~GeV$^2/c^2$, 
but with a similar slope. The partonic calculation of Kivel and Vanderhaeghen (2009) \cite{Kivel:2009eg}, particularly the calculation labelled
COZ is above the phenomenological estimate at $Q^2=2.5$~GeV$^2/c^2$, but with a shallower slope. The prediction of Kuraev et al.\ (2008)
\cite{Kuraev:2007dn}, whose authors criticize the TPE hypothesis, is below the phenomenological estimate. 

When viewed as a function of $Q^2$, the phenomenological prediction continues to grow, which reflects the fact that experimental
form factor data indicate that the discrepancy does not abate at higher $Q^2$. Neither the hadronic nor partonic calculations
continue to grow with $Q^2$. 

A region of particular importance for new measurements is the region of $2<Q^2<5$~GeV$^2/c^2$. In this region, neither the hadronic calculations
nor the partonic calculations are expected to perform accurately. The best bet for constraining hard TPE in this range is experimental data.

\subsection{Previous Measurements}

Hard TPE can be probed experimentally through a number of observables. The positron-proton to electron-proton cross section ratio probes
the real part of the TPE amplitude. One can also probe the imaginary part of the TPE amplitude through measurements of single-spin asymmetries
in which the polarization vector is normal to the scattering plane. In this section, we review only the former, since the leading
TPE correction to the unpolarized cross section (i.e., the effect on extractions of proton form factors) comes from the real part of TPE. 

\begin{figure}
    \centering
    \includegraphics{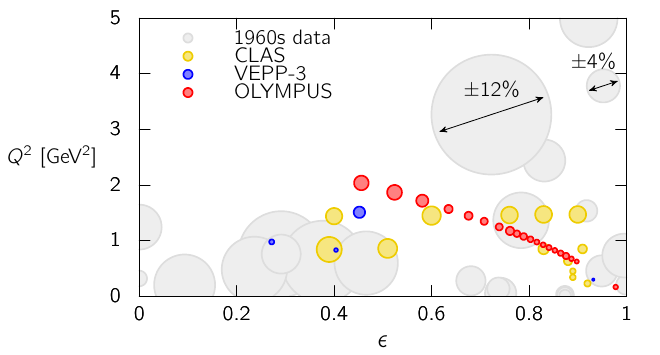}
    \caption{A kinematic map in the $\epsilon$-$Q^2$ plane of previous measurements of $R_{2\gamma}$. Each data
    point is represented by a circle whose diameter is proportional to its uncertainty. 
    Data from the 1960s~\cite{Yount:1962,Browman:1965zz,Mar:1968qd,Anderson:1966,Cassiday:1967,Anderson:1968zzc,Bartel:1967dsa,Bouquet:1968yqa}
    are shown as light gray circles. Data from VEPP-3~\cite{Rachek:2014fam} are shown as blue circles, data from CLAS~\cite{CLAS:2016fvy} are shown as yellow circles, while data from OLYMPUS~\cite{OLYMPUS:2016gso} are shown as red circles. 
    }    
    \label{fig:prev_data_kin}
\end{figure}

Comparisons of the positron-proton and electron-proton elastic cross sections to look for signs of TPE were carried out at a number of laboratories in the 1960s, including at Stanford~\cite{Yount:1962,Browman:1965zz,Mar:1968qd}, Cornell~\cite{Anderson:1966,Cassiday:1967,Anderson:1968zzc}, DESY~\cite{Bartel:1967dsa}, and Orsay~\cite{Bouquet:1968yqa}. 
The kinematics and uncertainties of these data are shown in Fig.~\ref{fig:prev_data_kin} as gray circles. 
From Figs.~\ref{fig:tpe_pred_QSq250} and \ref{fig:tpe_pred_eps742}, the best kinematics for observing a hard TPE signature are at low-$\epsilon$, i.e., backward scattering angles, where the effect size needs to be larger to produce the necessary bias to Rosenbluth plots, and at large $Q^2$, where the discrepancy is larger.
The data from the 1960s, by contrast, are predominantly at either high-$\epsilon$ or low $Q^2$. Furthermore, the typical uncertainties 
are of order 5--10\%, i.e., too large to resolve the expected effect size. Unsurprisingly, the 1960s experiments did not observe any significant hard TPE effects.

\begin{figure}[p]
    \centering
    \includegraphics{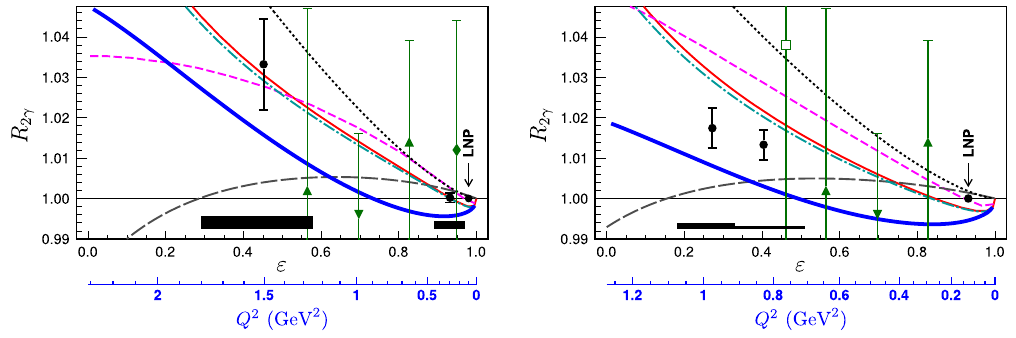}  
    \caption{Results from the two-photon exchange experiment conducted at the VEPP-3 storage ring at the Budker Institute, Novosibirsk, Russia, taken from Ref.~\cite{Rachek:2014fam}. The left panel shows the results for a beam energy of 1.6~GeV, while the right panel shows the results for a beam energy of 1~GeV. The filled black circles show the VEPP-3 results, while the other data points come from previous measurements: open squares from Ref.~\cite{Browman:1965zz}, closed upward-pointing triangles from Ref.~\cite{Anderson:1968zzc}, closed downward-pointing triangles from Ref.~\cite{Anderson:1966}, and closed diamonds from Ref.~\cite{Bartel:1967dsa}. Theoretical and phenomenological predictions are shown as curves for comparison: the dash-dotted teal curve is from Ref.~\cite{Borisyuk:2008es}, the red curve is from Ref.~\cite{Blunden:2005ew}, the thick blue curve is from Ref.~\cite{A1:2013fsc}, the gray dashed curve is from Ref.~\cite{Tomasi-Gustafsson:2009cfi}, the magenta dashed curve is from Ref.~\cite{Arrington:2004is}, and the gray dotted curve is from Ref.~\cite{Qattan:2011ke}.
    }
    \label{fig:vepp3}
\end{figure}

\begin{figure}[p]
    \centering
    \includegraphics[width=0.49\textwidth]{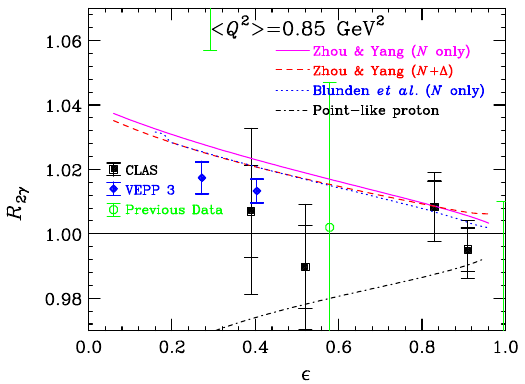}
        \includegraphics[width=0.49\textwidth]{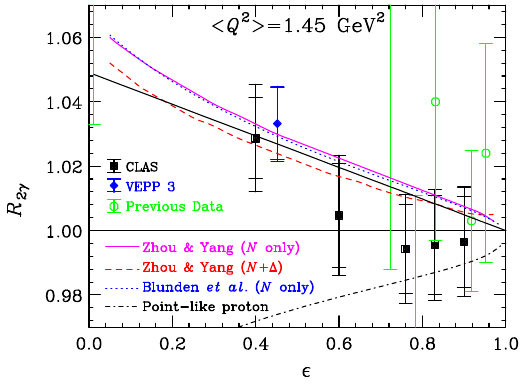}   
    \caption{Results from the two-photon exchange experiment performed with CLAS, taken from Ref.~\cite{CLAS:2016fvy}. The left and right panels show the $Q^2=0.85$~GeV$^2$ and $Q^2=1.45$~GeV$^2$ settings, respectively. The VEPP-3 data (blue points) come from Ref.~\cite{Rachek:2014fam} (Fig.~\ref{fig:vepp3}), while the previous data (green points) come from Refs.~\cite{Yount:1962,Browman:1965zz,Mar:1968qd,Anderson:1966,Cassiday:1967,Anderson:1968zzc,Bartel:1967dsa,Bouquet:1968yqa}. 
    The predictions labeled ``Zhou \& Yang'' come from Ref.~\cite{Zhou:2014xka}, while the prediction of Blunden et al.\ comes from Ref.~\cite{Blunden:2005ew}, and the ``Point-like proton'' prediction comes from Ref.~\cite{Arrington:2011dn}
}
    \label{fig:clas6}
\end{figure}

The emergence of the form factor discrepancy and the inadequacy of previous $R_{2\gamma}$ data motivated three new experiments to measure hard TPE at the percent level. An experiment at the VEPP-3 storage ring, at the Budker Institute in Novosibirsk, Russia, used alternating beams of electrons and positrons, scattering from a gaseous hydrogen target, and measured scattered leptons and recoiling protons in non-magnetic detectors~\cite{Rachek:2014fam}. An experiment was conducted in Hall B at Jefferson Lab, as part of the eg5 run period, which in the absence of a positron beam used a tertiary beam of $e^- e^+$ pairs with a broad energy range and the CLAS detector~\cite{CLAS:2013mza,CLAS:2014xso,CLAS:2016fvy}. Finally, the OLYMPUS Experiment~\cite{OLYMPUS:2013lem,OLYMPUS:2016gso}, conducted at the DORIS storage ring at DESY, in Hamburg, Germany, used alternating electron and positron beams and a gaseous hydrogen target, but, unlike at Novosibirsk, reconstructed events with a toroidal magnetic spectrometer. The fact that none of the three experiments used the same experimental technique gave each set of results different sensitivities to systematic effects. The kinematics of the data from these new experiments is also shown in Fig.~\ref{fig:prev_data_kin}. 

The results of the VEPP-3 TPE experiment are shown in Fig.~\ref{fig:vepp3}. The experiment had runs with a 1.6~GeV and 1.0~GeV beam energy. The VEPP-3 data show the largest effect of the three contemporary experiments. $R_{2\gamma}$ grows with decreasing $\epsilon$ and increasing $Q^2$, as would be expected given the form factor discrepancy, and reaches a few percent above unity. Because VEPP-3 had non-magnetic detectors, the experiment lacked a way to momentum analyze out-going particles. Elastic scattering events were separated from inelastic background through calorimetric energy measurements and angular correlations. The experiment also lacked an independent way to determine the relative $e^+$ and $e^-$ luminosities. Instead the data are reported relative to $R_{2\gamma}$ at a ``luminosity normalization point'' (LNP). Therefore the results should be interpreted as $R_{2\gamma}/R_{2\gamma}^\text{LNP}$. When comparing to theoretical predictions, the VEPP-3 data should be rescaled to the theory prediction for $R_{2\gamma}$ at the LNP.

The results of the CLAS TPE experiment are shown in Fig.~\ref{fig:clas6}. Because of the wide energy spread of the $e^+e^-$ pair beam, the experiment published results in two-dimensional $\epsilon$-$Q^2$ bins. The incoming beam particle energy was reconstructed from the kinematics of the outgoing lepton and proton, while the lepton charge was determined from curvature direction in the CLAS toroidal magnetic field. 
Two different binning schemes were used: one in which the bin centers had approximately constant $Q^2$ and varied in $\epsilon$, and another in which the bin centers had approximately constant $\epsilon$ and varied in $Q^2$. 
The two sets of results are not statistically independent as the bins overlap. 
Because a $e^+e^-$ pair beam was used, there was no need for an independent luminosity normalization between the two lepton charges. 
From the CLAS results, the lowest $\epsilon$, highest $Q^2$ data point is approximately 3\% above unity ($\pm 1.4\%$), but all other data points
are consistent with no TPE effect, given the large uncertainties, which are primarily driven by statistics. 

\begin{figure}
    \centering
    \includegraphics{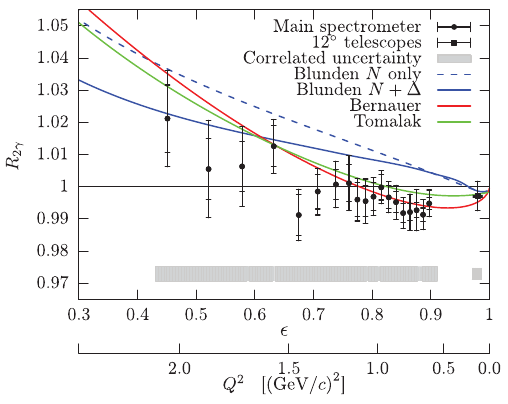}  
    \caption{Results from the OLYMPUS experiment, performed at DESY in Hamburg, Germany, taken from Ref.~\cite{OLYMPUS:2016gso}. }
    \label{fig:olympus}
\end{figure}

The results of the OLYMPUS experiment are shown in Fig.~\ref{fig:olympus}. OLYMPUS used a large-acceptance toroidal magnetic spectrometer, similar to CLAS, as well as a storage ring and gaseous target like the VEPP-3 experiment, and so shared similarities with both experiments. Data were collected in 2012--13 at DESY, and published in 2017, with a single fixed beam energy of 2~GeV. OLYMPUS determined the relative $e^+$ and $e^-$ luminosities in an entirely different way to the VEPP-3 experiment. OLYMPUS used a pair of forward calorimeters to compare rates of symmetric M\o ller/Bhabha scattering and forward elastic $e^\pm p$ scattering in each bunch crossing~\cite{Schmidt:2017jby}.  While OLYMPUS had the largest total data set in terms of integrated luminosity times acceptance, the experiment had to overcome a number of systematic hurdles, the most significant of which was unacceptable detector performance in the electrons-outbending toroid polarity setting. Both the OLYMPUS and CLAS experiments planned frequent reversals of all magnet polarities to suppress systematic differences between electron and positron detection. In OLYMPUS, with the detector only able to operate in one torus polarity, a great deal of effort needed to be expended to understand the small scale details of the detector efficiency over the entire acceptance. The OLYMPUS results show an $R_{2\gamma}$ that dips below 1 at high $\epsilon$, but then shows and upward trend at lower $\epsilon$ and higher $Q^2$. This dip is a feature of some phenomenological estimates~\cite{A1:2013fsc,Schmidt:2019vpr} (but not all~\cite{Chen:2007ac,Guttmann:2010au}), and is not present in the hadronic predictions of Blunden et al.~\cite{Blunden:2017nby,Ahmed:2020uso}.

\subsection{Current Status of Hard TPE}

The current state of knowledge of hard TPE is an uncomfortable one. There is a clear effect producing some kind of bias between determinations of the proton's form factors based on unpolarized cross sections, and those based on polarization observables. Hard TPE, as a neglected radiative correction is a natural candidate for the source of the effect. The effect is difficult to calculate, and the calculations we have come with model dependence. A recent round of experiments attempted to measure the hard TPE effect and found underwhelming results. This is in part because none of the experiments had the kinematic reach to probe hard TPE in kinematics where the proton form factor discrepancy is large. As seen in Fig.~\ref{fig:GE_GM}, the recent experiments were able to cover the region up to $Q^2=2$~GeV$^2/c^2$, whereas the discrepancy is large between 2 and 7~GeV$^2/c^2$ (or even higher, given the recent results from Hall A~\cite{Christy:2021snt}). Without data covering this region, we do not know whether hard TPE is truly the cause of the discrepancy, or if a different effect is in play (for example, the treatment of other radiative corrections, multi-photon emissions, etc.). 

In this introduction, we have discussed hard TPE in the context of the simplest reaction, elastic electron-proton scattering; however, radiative corrections are also critical for the interpretation of more complicated reactions, such as deeply virtual Compton scattering or deeply exclusive meson production. The effects of hard TPE may also produce bias in these reactions, and continue to cloud our understanding of nucleon structure.

The future positron facility planned at Jefferson Lab, combined with the high luminosity and large acceptance of the CLAS12 detector would open up the possibility of a high-statistics measurement of hard TPE covering a wide swath of kinematic space. Such an experiment would be a decisive test of whether or not hard TPE is the cause of the form factor discrepancy, build a firm bridge between the kinematic regimes suited for partonic and hadronic calculations, and provide valuable constraints for our progressing understanding of radiative corrections. 

This proposal is developed from Section 5.4 Projected measurements at CLAS12 (pp. 40--21) of letter of intent submitted to PAC 46 in 2018 (\href{https://www.jlab.org/exp_prog/proposals/18/LOI12-18-004.pdf}{LOI 12-18-004} \cite{positronLOI}). That letter covered a wide range of physics that would be made possible by a positron beam at CEBAF. It was developed by the Jefferson Lab Positron Working Group into a white paper, published as a topical issue of the European Physical Journal A~\cite{Alamanos:2022wwn}. A preliminary concept for the experiment proposed here was published in that issue~\cite{Bernauer:2021vbn}. A summary of the general physics case for a positron beam at Jefferson Lab, as well as an overview of how it can be implemented can be found in Ref.~\cite{Accardi:2020swt}.

\section{Experimental Concept}
\subsection{Overview}

Our proposed experiment will measure the cross section ratio of positron-proton elastic scattering relative to electron proton elastic scattering,
$R_{2\gamma}\equiv \frac{\sigma_{e^+p}}{\sigma_e^-p}$. Alternating positron and electron beams will be scattered from a liquid hydrogen target with
both the scattered lepton and recoiling proton detected in coincidence in the CLAS12 spectrometer in Hall B. This experiment will require the future positron beam capabilities planned for CEBAF.

The motivation for the experiment calls for measurements over a wide phase space in both $Q^2$ and $\epsilon$, focusing on $2 \lessapprox Q^2 \lessapprox 7$~GeV$^2/c^2$ at low $\epsilon$. For this reason, three different beam energies will be used, and CLAS12 will be configured for the detection of both forward and backward outgoing leptons. 

\begin{table}[htpb]
    \centering
    \caption{Comparison of $R_{2\gamma}$ Measurements}
    \begin{tabular}{c c c c | c}
    \hline
    \hline
         & VEPP-3 & CLAS & OLYMPUS & CLAS12 \\
         \hline
        Azimuthal Acceptance & 33\% & 50--85\% & 15\% & $>90\%$ \\
        Luminosity [cm$^{-2}$s$^{-1}$] & $10^{32}$ & $<5\cdot 10^{31}$ & $2\cdot 10^{33}$ & $10^{35}$ \\
        Beam Energy [GeV]& 1.0, 1.6 & variable & 2.0 & 2.2, 4.4, 6.6 \\
        \hline
    \end{tabular}
    \label{tab:r2g_exp_compare}
\end{table}

As shown in Fig.~\ref{fig:GE_GM}, the capabilities of CLAS12 and the planned capabilities of the CEBAF positron beam will lead to a major increase in kinematic reach compared to previous $R_{2\gamma}$ measurements. This can be seen from the experimental parameters given in Table.~\ref{tab:r2g_exp_compare}. The CLAS12 can run with a far higher luminosity than any of the previous experiments were able to achieve. In addition, CEBAF can deliver a much larger range of beam energies. This will allow a significantly higher precision measurement in overlapping kinematics and make a wide range of new kinematic territory accessible.

\subsection{Beams}

This experiment requires the future positron beam capabilities planned for Jefferson Lab. In the experiment, we plan to alternate between electron and positron beams, i.e., measuring with electrons in the same experiment and in the same manner as with the positrons, in order to minimize differences in running conditions of the two beam species. Rather than using the standard electron source, our experiment will use the positron production target, but with the polarity of the capture system reversed, so as to capture and re-accelerate electrons with a beam profile as similar as possible to the positrons. The goal is not to get the cleanest possible electron beam, but rather to get the positron and electron beams to be as similar as possible to minimize systematic differences between them. 

To minimize sensitivities to time-dependent performance changes in the detectors, we plan to interleave electron and positron running, which is the major reason we are strongly advocating for the ability to switch between electron and positron modes rapidly. For instance, in OLYMPUS, the transition between beam species was accomplished in 1--2 hours, and was typically performed once per day. While the eventual capabilities of the CEBAF positron beam are far from finalized, we envision here that the transition between beam species could be accomplished within one 8 hour shift, and that we would switch between species once per week, ideally to overlap with an RF recovery day.

Our experiment will use standard 1-pass, 2-pass, and 3-pass energies, i.e., 2.2, 4.4, and 6.6~GeV respectively, with the majority of the time being used for 3-pass running. As the goal of the experiment is to cover a wide two-dimensional phase space, the exact energy values that the accelerator can deliver are not critical. 

The beam will be scattered from a liquid hydrogen target in Hall B. For the standard CLAS12 5~cm liquid cell, a luminosity of $10^{35}$~cm$^{-2}$s$^{-1}$ would be a achieved at a beam current of approximately 60~nA. This is well within the planned capabilities of the CEBAF positron beam, described in the memo produced for PAC51 by the Positron Working Group~\cite{PWG:memo}. The current estimate calls for a source that can deliver up to 1~$\mu$A of unpolarized positrons. 

\subsection{Detector System}

\begin{figure}
    \centering
    \includegraphics[width=0.45\textwidth]{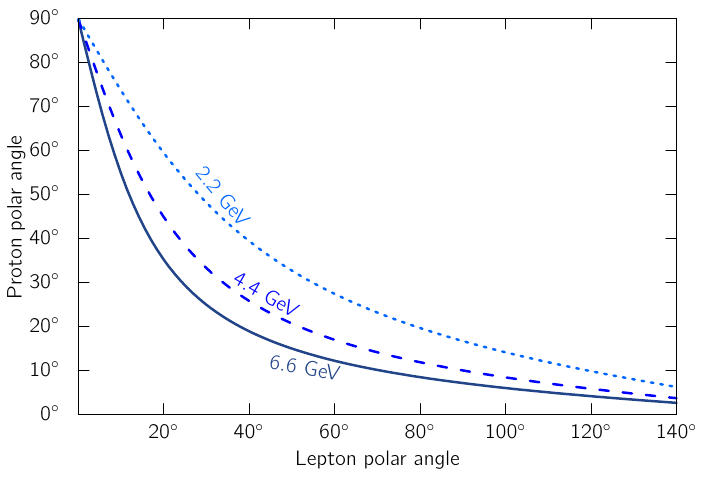}\hspace{0.05\textwidth}
    \includegraphics[width=0.45\textwidth]{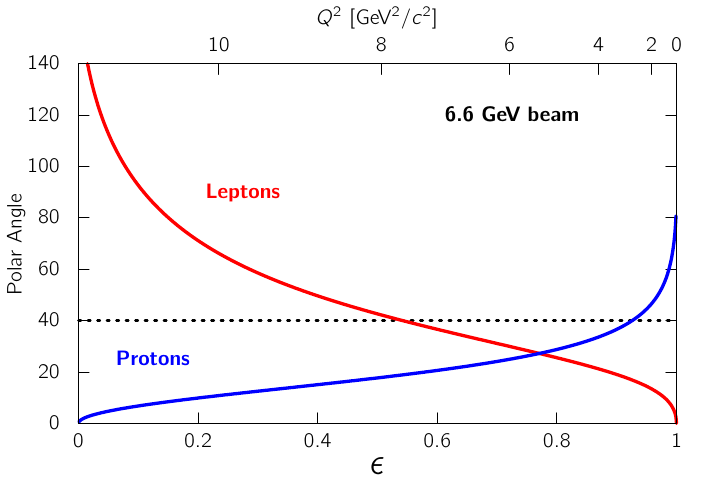}\\
            \includegraphics[width=0.45\textwidth]{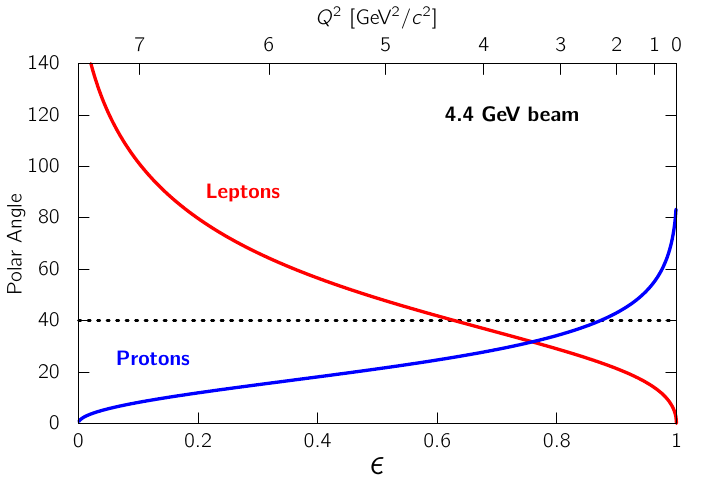}\hspace{0.05\textwidth}
    \includegraphics[width=0.45\textwidth]{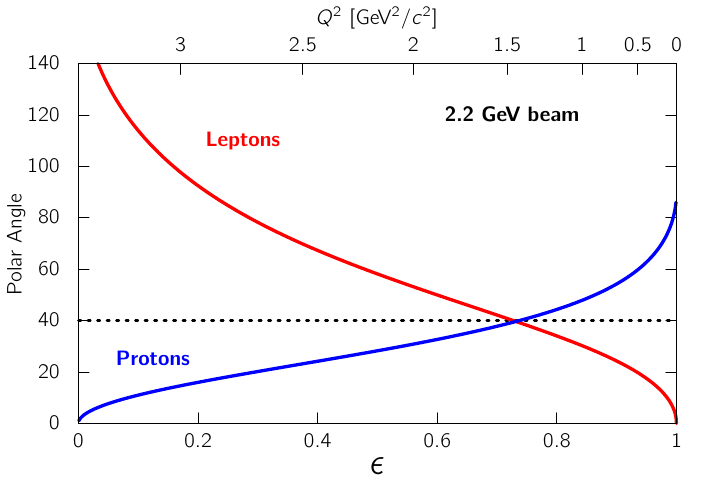}
    \caption{Upper left: The lepton angle/proton angle correlations for elastic scattering.
    Upper right: the correspondance between lepton angle, proton angle, $\epsilon$ and $Q^2$ for a 6.6~GeV beam energy.
    The dashed line at $\theta=40^\circ$ corresponds to the approximate boundary between the forward and central detectors.
    Lower left and right: the same correspondance for 4.4 and 2.2 GeV beam energies.}
    \label{fig:angle_angle}
\end{figure}

Elastic scattering events will be reconstructed by the coincident detection of both the scattered lepton and the recoiling proton in CLAS12. The advantage of coincident detection is the several degrees of over-constraint of the kinematics, allowing a high degree of separation between elastic signal and inelastic background. In particular, we can make use of angular correlations, which are measured with better resolution than momenta, to select elastic events. The relationships between lepton angle and proton angle for the three different beam energies are shown in the upper left panel of Fig.~\ref{fig:angle_angle}. The mapping between angles and $\epsilon$ and $Q^2$ are shown for a 6.6~GeV beam in the upper right panel, for a 4.4 GeV beam in the lower left panel, and for a 2.2 GeV beam in the lower right panel.

Being able to make use of over-constrained kinematics is important because, as can be seen in Fig.~\ref{fig:angle_angle}, for nearly all kinematics, one of the two outgoing particles will be detected in the CLAS12 central detector, where the momentum resolution is not as favorable, and where particle identification is more challenging. Whereas the momentum resolution in the forward detector is approximately 0.7\%, the typical momentum resolution in the central detector is about 3\%. The central detector has no Cherenkov detectors and no electromagnetic shower detectors, and the short tracking path length restricts the utility of momentum-timing correlations to perform particle ID. For a coincident measurement, good recontruction and particle ID can be performed on the forward particle, and the strong constraints of track coplanarity and polar-angle correlations can be used to identify the central particle and reject background. No additional hardware will be needed to supplement the standard CLAS12 central detectors. 

\begin{figure}
    \centering
    \includegraphics[width=0.45\textwidth]{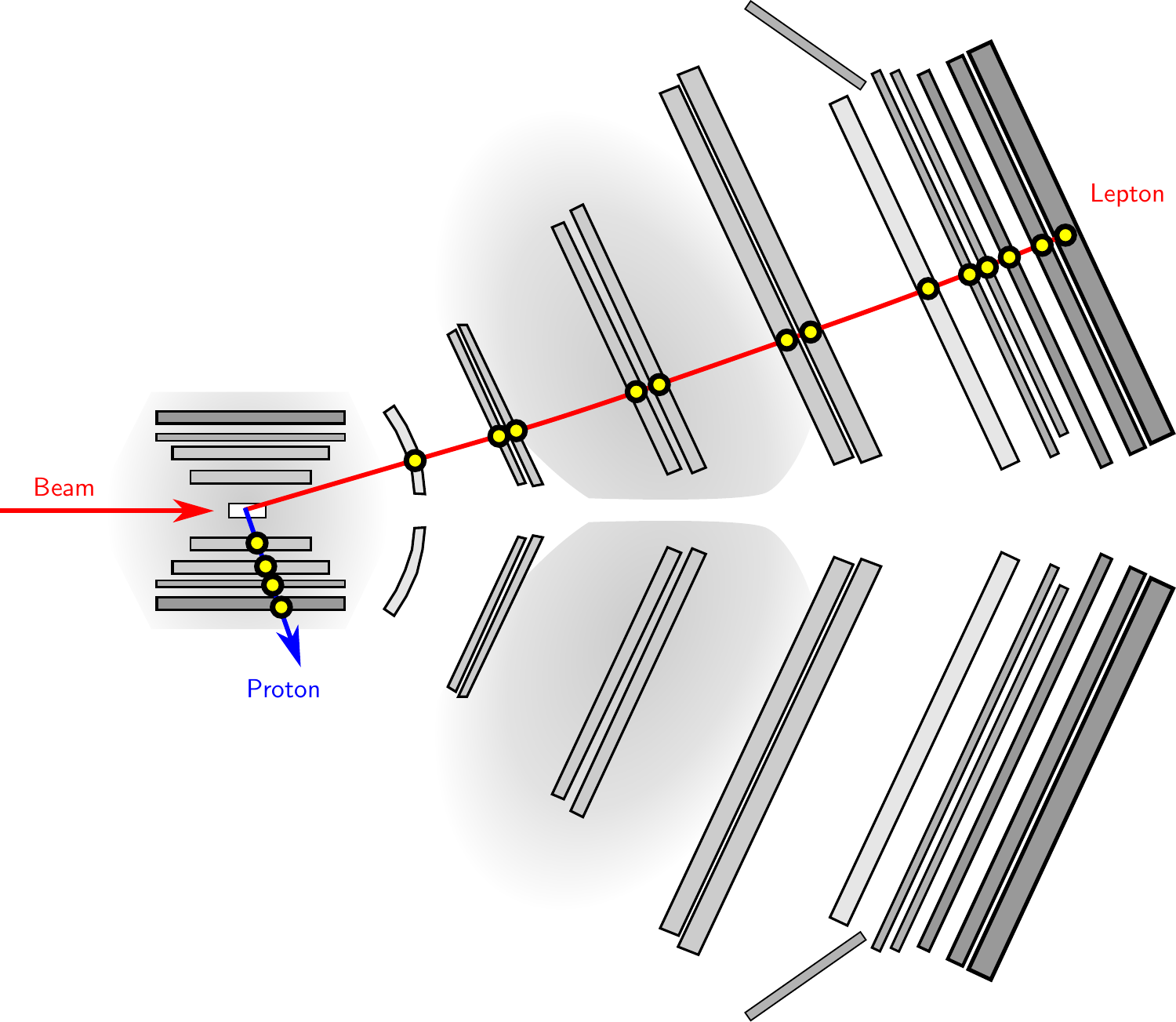}\hspace{0.05\textwidth}
    \includegraphics[width=0.45\textwidth]{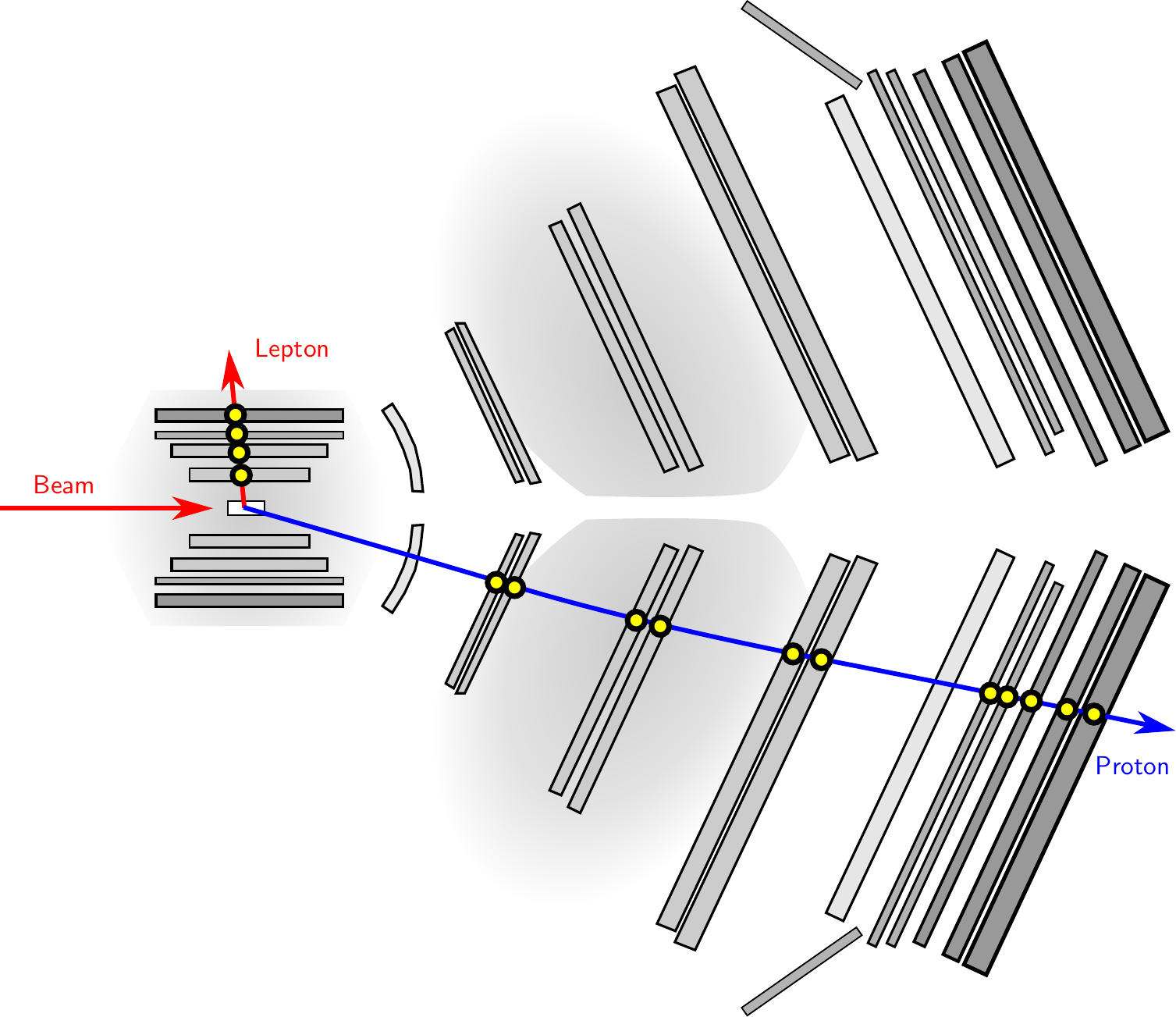}
    \caption{Illustrations of two example elastic events relative to a cross section of the CLAS12 detector
    (not to scale). The event on the left corresponds to a low-$Q^2$, high-$\epsilon$ event, with the forward
    torus in the beam-charge-outbending polarity configuration. The event on the right corresponds to a 
    high-$Q^2$, low-$\epsilon$ event, with the forward torus in the positive particles in-bending polarity configuration.}
    \label{fig:clas12_events}
\end{figure}

The geometry of the CLAS12 spectrometer leads to two general topologies for elastic events of interest. 
These are shown in Fig.~\ref{fig:clas12_events}. 
At low $Q^2$ and high $\epsilon$, the lepton will traverse the forward detector, while the proton will pass through the central detector. 
At high $Q^2$ and low $\epsilon$, the proton will recoil through the forward detector, while the lepton scatter through the central detector. (There may also be a small kinematic region where both particles go forward, as can be seen from the upper right panel of Fig.~\ref{fig:angle_angle}.) As will be discussed, the two topologies have different triggering/read-out requirements. The high-$Q^2$ rate will be much lower than the low-$Q^2$ rate.

\begin{figure}
    \centering
    \includegraphics{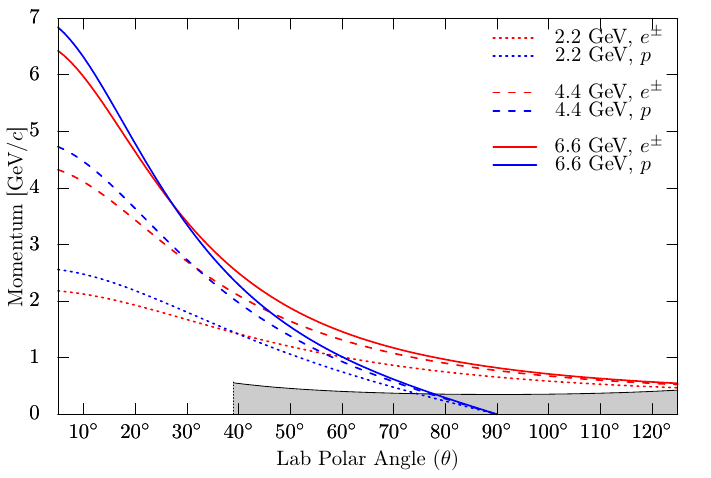}
    \caption{Particle momenta as a function of polar angle for the three different beam energies. The shaded area at the bottom indicates the region excluded by the minimum transverse momentum threshold for detection in the CLAS12 central detector ($p_T > $ 350 MeV/$c$).}
    \label{fig:mom_v_theta}
\end{figure}

Particle momenta are fixed as a function of scattering angle in elastic scattering. The lepton and proton momenta for the three different beam energies are shown in Fig.~\ref{fig:mom_v_theta}. Except at the lowest $Q^2$ values, the proton momenta will be above the central detector momentum threshold of approximately $p_T > 350$~MeV$/c$. Leptons should also be detectable in the central detector over the full angular range. 

\subsection{Magnet Polarity Reversals}

Magnetic fields will bend electrons and positrons with identical initial momentum vectors to different regions of the CLAS12 spectrometer. Variation in efficiency across the spectrometer can risk producing a false asymmetry between electron scattering and positron scattering. This will be a major systematic concern for this experiment, as it was for the prior round of TPE experiments. This potential false asymmetry can be suppressed by collecting data with both magnetic field polarities. $R_{2\gamma}$ can be constructed by taking the geometric mean of cross sections in both settings:
\begin{equation}
    R_{2\gamma} = \sqrt{\left(  \frac{\sigma_{e^+p}}{\sigma_{e^-p}}\right) _\uparrow \cdot \left( \frac{\sigma_{e^+p}}{\sigma_{e^-p}}\right)_\downarrow}.
    \label{eq:double_ratio}
\end{equation}
With magnetic fields reversed, inefficient areas for electrons will become inefficient areas for positrons, allowing a partial cancellation of effects.
This cancellation is not exact for a number of reasons, but it does help to significantly reduce the sensitivity of $R_{2\gamma}$ to any efficiency effects.

CLAS12 has two magnets, which can each run in either polarity setting. Particles being detected in the central detector are sensitive to the polarity of the solenoid. Particles traversing the forward detector are affected by both the solenoid and toroid polarity. The experiment will therefore use four different magnetic field configurations, allowing us to construct the observable as:
\begin{equation}
    R_{2\gamma} = \left[ \left(  \frac{\sigma_{e^+p}}{\sigma_{e^-p}}\right) _{\uparrow\uparrow}
    \cdot \left( \frac{\sigma_{e^+p}}{\sigma_{e^-p}}\right)_{\uparrow\downarrow}
    \cdot \left( \frac{\sigma_{e^+p}}{\sigma_{e^-p}}\right)_{\downarrow\uparrow}
    \cdot \left( \frac{\sigma_{e^+p}}{\sigma_{e^-p}}\right)_{\downarrow\downarrow}    \right]^{1/4}.
    \label{eq:quad_ratio}
\end{equation}

\subsection{Read-out}

The most significant change to standard CLAS12 running that this experiment will require is in the way CLAS12 is triggered. The standard CLAS12 trigger requires the detection of an energetic electron in the forward detector. For the majority of the kinematics of interest, the outgoing lepton will pass through the central detector, rather than the forward detector. A new triggering scheme will be needed, one that relies on the detection of a high-momentum hadron in the forward detector. 

A forward hadron trigger does pose challenges. CLAS12 Run Group K observed that a secondary trigger based on forward charged tracks had a rate of 420 kHz at beam energy and luminosity conditions similar to what we propose. This between a facto or 4--5$\times$ higher than the 100 kHz rate that should be feasible with the planned CLAS12 high-luminosity upgrade. Fortunately, there are several additional handles that can be used to achieve a $5\times$ rate reduction:
\begin{itemize}
    \item CTOF/CND coincidence: low-$\epsilon$ elastic events will produce a coincident hit in the central time-of-flight detector and/or the central neutron detector. 
    \item CVT coincidence: the scattered lepton will produce a track in the Central Vertex Tracker. A reconstructed track is not necessary; just the requirement of hits could be used as a trigger input. The forward detector frequently makes use of ``DC roads,'' hit information from the forward drift chambers, to aid in triggering. A similar system of ``CVT roads'' could be implemented.
    \item Kinematic correlations between forward and central detector tracks: because elastic events have very specific kinematics, this information could be exploited in the trigger. For example, the forward hadron and central lepton coincidence could be made to satisfy a co-planarity requirement, as well as a polar-angle correlation requirement.
    \item Cherenkov veto: the HTCC can be used to veto electrons and positrons scattering at forward angles. This will primarily help at lower beam energies, where electromagnetic scattering cross sections are higher.
\end{itemize}
The optimal solution will depend in great deal, on the capabilities of CLAS12 at the time of future positron operations. However, various trigger configurations could be tested with electrons as early as the next running period, to assess rates and feasibility.

Elastic scattering at low-$Q^2$/high-$\epsilon$ can be measured with the conventional CLAS12 electron trigger. For reasons associated with luminosity uncertainty, it would be preferable to measure the complete kinematic range simultaneously, rather than sequentially. Therefore, we propose to use a pre-scaled forward-electron trigger, in addition to the forward-hadron main trigger. The low-$Q^2$/high-$\epsilon$ elastic rate is so high that a substantial pre-scale factor could be used.

Many upcoming nuclear physics experiments plan to use a streaming readout paradigm, rather than triggered-read out. There has already been some R\&D into using streaming readout in CLAS12, specifically for the forward tagger~\cite{Ameli:2021agc}. While there are no current plans to adopt streaming readout for the entire CLAS12 detector, there also may be a long time before the Jefferson Lab positron program is executed, and the readout capabilities of CLAS12 might be substantially different by the time this experiment would be on the floor. It is worth mentioning that if streaming readout schemes were adopted, the issues of identifying and recording central-lepton/forward-proton elastic events would essentially become moot. By performing a basic track reconstruction, the tight kinematic correlations between lepton and proton tracks could be used to identify elastic events and reject background to a high degree.

\subsection{Luminosity}

Because the electron-scattering and positron-scattering data would be collected in different data sets, the relative integrated luminosity between the two data sets becomes a crucial normalization. The three recent TPE experiments handled this issue in different ways, with VEPP-3 choosing to normalize relative to rates of low-$Q^2$ elastic scattering, OLYMPUS using a pair of far-forward M\o ller/Bhabha calorimeters, and CLAS avoiding the problem by measuring with an $e^+e^-$ pair beam. Luminosity in Hall B has typically been estimated from a measurement of the integrated beam charge by the Faraday cup combined with an estimate of the target density. For reference, the most recently published absolute cross section measurements by CLAS (e.g.~Refs.~\cite{CLAS:2014fml,CLAS:2015uuo,CLAS:2017rgp}) claimed absolute luminosity uncertainties of 2--5\%. The Hall B beam dump is planned to be upgraded by the time of the JLab positron program, and it is anticipated that the most accurate current measurements will come from cavity beam current monitors (BCMs), which should have comparable or even better absolute performance.

However, our proposed experiment we propose does \emph{not} require an absolute determination of the luminosity; only the relative luminosity between $e^-$ and $e^+$ settings is needed, and this is known far-better than 2--5\%. The principle uncertainty will come from the charge determination rather than the target density. At these beam currents, we can expect very little target boiling effect. The target temperature and pressure will need to be monitored continuously, and any time-dependencies accounted for in analysis. For the BCMs, the relative response to positive and negative currents will need to be studied, once the system is fully implemented. 

Even if a substantial, uncontrollable systematic uncertainty is discovered in the performance of the BCMs, the experiment can still normalize to the elastic scattering rates at the lowest $Q^2$ data point at each beam energy. This is less preferable because the normalization would become dependent on the (unknown) TPE effect at the normalization point. However, this was the approach used in the experiment at VEPP-3~\cite{Rachek:2014fam}. TPE is predicted to be a small effect at forward angles, with a similarly small effect on the normalization.

\subsection{Strategy for Testing the TPE Hypothesis}

This experiment will add several dozen new measurements of $R_{2\gamma}$ over a wide range of $\epsilon,Q^2$ space. These data will provide valuable constraints of theoretical calculations and phenomenological estimates over that space. We are not planning to match specific kinematics of existing or upcoming form factor measurements; we do not feel that the effort is justified, especially since this will depend on the exact beam energies available at the time of running. New theoretical calculations and phenomenological reanalyses will take time. Nevertheless, our data can be immediately used to assess the TPE hypothesis by comparing to phenomenological estimates. This approach promptly used after the most recent round of TPE experiments~\cite{Afanasev:2017gsk,Schmidt:2019vpr}.

\section{Radiative corrections}

Radiative corrections are another critical aspect of the proposed experiment. Elastic and quasielastic electron scattering experiments commonly use the ``peaking approximation'' approach developed by the NE-18 experiment~\cite{Ent:2001hm}, and implemented in a number of common Monte Carlo event generators, including SIMC (Hall A/C physics Monte Carlo). This method would be inadequate for this experiment, and a more sophisticated approach will be necessary. Fortunately, there have been a number of important developments in radiative correction theory, spurred largely by the last round of TPE experiments. 

As outlined above, hard two photon exchange is defined as the difference of all TPE to the ``soft" part that is included in the radiative corrections that are applied to the data. It is not possible to ignore soft TPE entirely because it has infrared-divergent behavior, and this divergence is necessary to cancel the infrared divergence in other diagrams. However, since different radiative correction prescriptions (e.g.~\cite{Mo:1968cg} vs.\ \cite{Maximon:2000hk}) differ in their treatment of soft TPE, the definition of hard TPE becomes prescription-dependent. When reporting a measurement of $R_{2\gamma}$, one must also report which definition of hard TPE was used, based on the soft TPE correction applied. OLYMPUS, for example, chose to report results for four different radiative correction prescriptions~\cite{OLYMPUS:2016gso}. 

The standard modern approach to apply radiative corrections is through the use of a Monte Carlo event generator that includes the effects of higher-order diagrams. Post-hoc applications of radiative corrections as multiplicative factors are not precise enough, especially when both particles are detected. Such an event generator ideally produces the full event topology, i.e., does not integrate over the proton kinematics, includes all interference terms, and does not use any kinematic short cuts like the peaking approximation. Typical implementations of generators only include NLO, or in some cases, NNLO corrections. The effects of any higher-order corrections are often neglected or approximated via exponentiation, i.e.,
\begin{equation}
    \left(\frac{d\sigma}{d\Omega}\right)_\text{exp.} = \left(\frac{d\sigma}{d\Omega}\right)_\text{Born} \cdot (1 + \delta) \rightarrow \left(\frac{d\sigma}{d\Omega}\right)_\text{Born} e^\delta,
\end{equation}
which is only correct in the limit of no radiated photon energy. However, a real measurement will have a finite energy resolution, i.e.,\ will always accept some soft bremsstrahlung events as elastic, and indeed, the energy cut is often defined to be larger than the typical energy resolution in order to increase counts and to minimize systematic effects from the cut definition.  Up to a given point, a larger accepted energy range will produce an apparently smaller radiative correction. However, this should not be misinterpreted as leading to smaller uncertainties, as it essentially results from the cancellation of two correction terms with independent uncertainties. Indeed, a large, but theoretically better described correction for elastic events is then preferable to a smaller correction with larger theoretical uncertainties. 

\begin{figure}
\centering
    \includegraphics{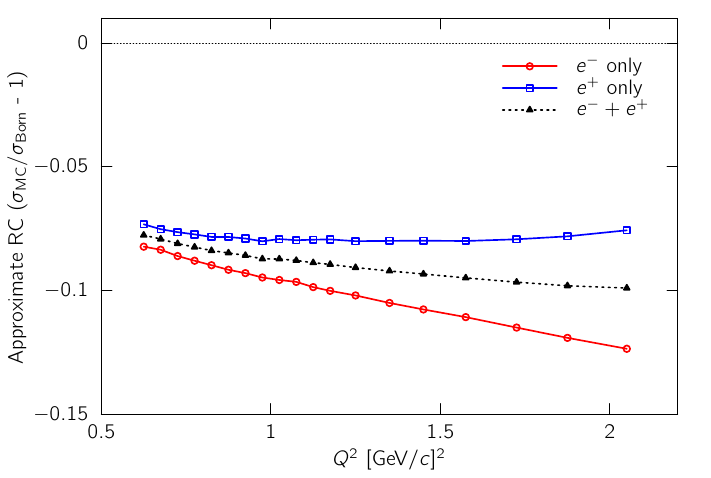}
    \caption{Estimate of the total radiative corrections for OLYMPUS cross section data (Fig.\ from \cite{OLYMPUS:2020dgl}). The charge-even components (black line) are dominating the charge-odd components (difference to colored lines). They change only minimally over the accepted kinematical range.}
    \label{fig:olympusRC}
\end{figure}

\begin{figure}
    \centering
    \includegraphics{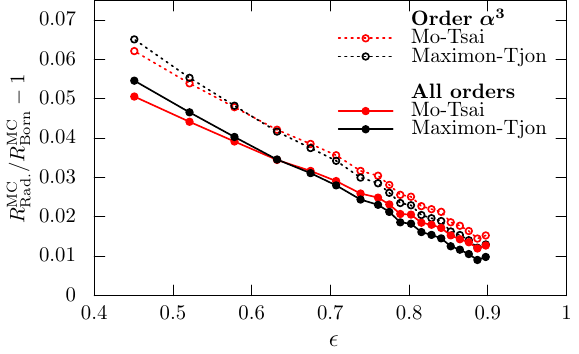}
    \caption{Estimate of the lepton charge-odd corrections on $R_{2\gamma}$, as applied in OLYMPUS (Fig. from \cite{OLYMPUS:2016gso}). As can be seen, the charge-odd corrections are sizable, a factor of a few larger than the extracted TPE contribution. Also shown is the dependence on used radiative corrections description  (red/black), and on the inclusion of higher order terms (solid/dashed). Overall, the dependence of the charge-odd terms on $\epsilon$ is strong. See text for details.}
    \label{fig:chargeoddolympus}
\end{figure}

Charge-even corrections (i.e., identical for electron- and positron-scattering) often dominate the corrections for cross section measurements (see Fig.\ \ref{fig:olympusRC}), and were therefore the focus of many earlier generator implementations. However, charge-odd terms, for example from the interference of bremsstrahlung off the lepton vs.\ proton, have not always been fully reproduced. For a TPE measurement, on the other hand, charge-even corrections only lead to a small dilution of the TPE effect on the ratio. Charge-odd terms, however, affect the ratio in the same manner as TPE; it is essential to model them correctly.  The size of these charge-odd corrections are substantial, see Fig.\ \ref{fig:chargeoddolympus} for values from OLYMPUS. The $y$-axis shows the simulated $\sigma_{e^+p}/\sigma_{e^-p}$ ratio, including standard radiative effects (i.e., only taking into account soft TPE), minus one. The deviation from zero indicates the size of the charge-odd correction. The results are shown for four different models of the radiative effects, in order to show their impact. The red points use the Mo and Tsai prescription~\cite{Mo:1968cg}, while the black points use the Maximon and Tjon prescription~\cite{Maximon:2000hk}. The open points show the results for when the prescription models the radiative effects to next-to-leading order, i.e., to the order of $\alpha^3$, where $\alpha$ is the fine structure constant. The filled points show the results using an exponentiated prescription to estimate the radiative effects to all orders. When measuring an asymmetry that is expected to only be a few percent, accuracy in these corrections is critical. It should be noted that the size of these corrections depends on the accepted energy range for bremsstrahlung events, which will be different for different experiments.

Various radiative event generators at the NLO level currently exist, including ESEPP \cite{Gramolin:2014pva} and the generator used for OLYMPUS. NNLO generators are becoming available, and we expect to use these for the final analysis. The performance of a generator can be checked in the experiment by comparing the description of the ``radiative tail'' as a function of the cut-off energy in MC and data - the produced ratio should be independent of the choice of cut-off energy. Current generators pass this test at least for lower beam energies \cite{Mihovilovic:2016rkr}.

Improved radiative corrections are a big focus of the community, as they are important for many precision measurements, from proton radius to EIC physics.  A current review of the state of the field can be found in the upcoming whitepaper \cite{radcorwp}.

\section{Systematic Uncertainties}

There are a number of physical and experimental factors that can produce a false asymmetry between electrons and positrons.
The systematic uncertainties in this experiment reflect our knowledge of these factors and the degree to which they can be quantified and corrected.
These factors include the use of magnetic fields (in combination with position- and time-dependent detector inefficiencies), radiative corrections, inelastic background, and the relative luminosity between beam species. While we address each in the subsequent sections, we begin with a review 
of what was learned about reducing systematic uncertainties from the previous generation of TPE experiments.

\subsection{Lessons learned from earlier experiments}

 Conceptually, the proposed work is closest to the OLYMPUS experiment, detecting both lepton and proton over a wide range of angles with a magnetic spectrometer, with mono-energetic beams. Therefore, we draw primarily from the experience from OLYMPUS. 

\subsubsection*{Define kinematics based on the proton}

The major source of point-to-point systematic uncertainties is the use of magnetic fields, which bend the electrons and positrons in different directions.
The use of magnetic fields for the momentum analysis is helpful for reducing inelastic background, but their use requires time and effort to understand the detector acceptance for in-bending and out-bending particles. Even worse, additional systematics are immediately introduced if the reconstructed kinematics depend in any way on the lepton bend direction, through effects such as differing resolutions or track-reconstruction biases. Because of the strong angular dependence of the scattering cross sections, even small systematic differences might affect the extracted ratio strongly enough to invalidate the result. Fortunately, this potential bias can be strongly suppressed by determining the event kinematics purely from the reconstructed proton. Lepton information can still be used to identify scattering events and reduce background, but $Q^2$ and $\epsilon$ are best defined in terms of the reconstructed proton, and thus are identical for electron and positron modes, regardless of the magnetic fields.  Possible biases are thus unchanged and largely cancel when taking a $\sigma_{e^+p}/\sigma_{e^-p}$ ratio. 

\subsubsection*{Keep lepton kinematic cuts wide}

For similar reasons, cuts on lepton information should be kept wide, so that any slight resolution differences due to bending direction have a minimal impact on reconstruction efficiency. In OLYMPUS, it was found that even if with kinematic cuts on the order of $\pm 5$--$8 \sigma$, the over-constrained kinematics made it very easy to isolate the elastic signal from background. When kinematic cuts were tightened, the impact of slight deviations between the data and Monte Carlo were enhanced, as seen when comparing different sectors. 

\subsubsection*{Use different sectors to perform unbiased studies of acceptance}

The most accurate way to study the detector acceptance is through a highly detailed Monte Carlo simulation in which the detector response is modeled as accurately as possible. This accuracy needs to be assessed in a way that is blind to physics. One simple and important way to do this is to take double ratios of data to simulation between two independent groups of data. A natural (but not unique) grouping for CLAS12 would be the forward-detector sector. 
By binning the data sector by sector, one can form a super-ratio between different groups $i,j$, 
\begin{equation}
    r_{i,j}=\left.\frac{\sigma_\textrm{exp}}{\sigma_\textrm{MC}}\right|_i/\left.\frac{\sigma_\textrm{exp}}{\sigma_\textrm{MC}}\right|_j.
\end{equation}
Ideally, $r_{i,j}=1$, as all physics, including the chosen form factor model within the Monte Carlo event generator, will cancel in the ratio. The spread of $r_{i,j}$ around 1 for each bin then directly measures the quality of the agreement between MC and experiment.

One can further take quadruple ratios of this above double ratio for positrons relative to that for electrons, $\mathcal{R}_{i,j} = r_{i,j}^{e^+p}/r_{i,j}^{e^-p}$, for which deviations from unity can highlight if any observed data/MC discrepancies will produce biases in $R_{2\gamma}$. 
These deviations can be used to quantify the systematic uncertainty stemming from the accuracy of the Monte Carlo simulation.

\subsubsection*{Avoid reliance on a luminosity normalization point}

The earlier results, mainly those from OLYMPUS, suggest that $R_{2\gamma}$ dips below unity at high-$\epsilon$, a feature that is not predicted in hadronic theory calculations. This makes the high-$\epsilon$ region interesting for study as well as the low-$\epsilon$ region. But in order to study this feature, one needs a measurement of $R_{2\gamma}$ that is not measured relative to a high-$\epsilon$ luminosity normalization point. Of the three previous measurements, the VEPP-3 result is relative to $R_{2\gamma}$ at a high-$\epsilon$ luminosity normalization point, as they did not have good independent control of the luminosity. Because of this, the utility of the VEPP-3 data for testing the behavior of $R_{2\gamma}$ at large $\epsilon$ is drastically reduced. This underscores the importance of having an independent measurement of the luminosity.

\subsubsection*{Charge-averaged cross sections}

Sufficient understanding of the detector and control of systematics also allow for measurement of hard TPE. Instead, the charge average of the measured cross section is free of TPE, indeed free of all charge-odd correction, and can be used to determine the proton form factors with smaller dependence on theoretical corrections. OLYMPUS recently released the first data set of this kind \cite{OLYMPUS:2020dgl}. While it requires more work to reduce systematics that cancel in the ratio but not in the average, we believe it is worthwhile.

\subsection{Magnetic fields, efficiency, and detector performance stability}

The largest point-to-point systematic uncertainty will likely come from the use of magnetic fields, which will bend electrons and positrons in different directions. The effect on $R_{2\gamma}$ is non-trivial. For electrons and positrons with identical kinematics, the magnetic fields will bend the two to different regions of the detector. Non-uniformities in detection efficiency within the fiducial volume will affect electrons and positrons differently. In principle, this effect could be simulated with a sufficiently detailed Monte Carlo and corrected. In practice, even after using a Monte Carlo correction, there is a residual systematic uncertainty due to uncertainty in the accuracy of the Monte Carlo itself. 

This magnetic field systematic can be further suppressed by reversing the magnetic fields, collecting data with both polarities, and combining the data according to Eqs.~\ref{eq:double_ratio} or \ref{eq:quad_ratio}. The electron efficiency in one polarity will become the positron polarity in the other, and vice versa, allowing a cancellation. The proton efficiency will change under the field reversal as well, but will cancel across configurations of identical polarity in Eq.~\ref{eq:double_ratio}. 

While field reversals are a powerful way to suppress systematics, we stress that this cancellation is not perfect. One problem is that the detector performance can fluctuate over time. The different magnetic field settings must be run one at a time, and if the detector performance changes, then the efficiencies will not exactly cancel. Another problem is that different field settings can lead to significantly different regions of acceptance. In CLAS12, the solenoid magnet produces a slight azimuthal rotation to tracks prior to their reaching the forward detector. For positrons, this rotation will always be in the same direction as the protons, regardless of polarity. For electrons, this rotation will always be in the opposite direction. This, combined with gaps in azimuthal coverage in the forward detector due to the torus coils can lead to significant differences in the fiducial acceptance between polarity settings. A drastic solution is to only consider a subset of the data, i.e., events in which both the lepton and proton would have entered the fiducial acceptance in all species and field configurations. This can be relaxed as the detector acceptance and magnetic fields become better understood. 

For this experiment, we aim to produce a Monte Carlo simulation that models the time-dependent performance of the detector, i.e., simulates conditions for individual runs. As the Monte Carlo is tuned and improved, we can quantify its accuracy by comparing data-to-simulation double ratios across sectors. A point-to-point systematic uncertainty of $\approx 1\%$ on $R_{2\gamma}$ will be needed to study TPE at low-$\epsilon$. In order to study the ``dip'' observed in OLYMPUS at high-$\epsilon$, a systematic uncertainty of 0.5\% or better will be necessary. 

For context, OLYMPUS, which did not flip the polarity of its toroid, achieved a point-to-point systematic uncertainties ranging from 0.33\%--1.18\%~\cite{OLYMPUS:2016gso}. These uncertainties were achieved by detailed modelling of the time-dependent detector performance in a Monte Carlo simulation. The level of detail included modelling the time-dependent, spatially-dependent, and track-angle-dependent efficiency of every drift chamber wire individually~\cite{Henderson:thesis,Russell:thesis,Schmidt:thesis}. In our proposed experiment, magnetic field reversals will make the problem significantly less arduous. 

\subsection{Radiative Corrections}

As discussed above, most radiative effects are charge-even. While these corrections can be large, they cancel in the ratio. More critical is the interference of bremsstrahlung from lepton and proton. Here, a series of modeling choices can affect the resulting corrections, for example the treatment of off-shellness of the intermediate state, as well as the treatment of higher order corrections. Since the bremsstrahlung corrections depend on the cut-off energy, the radiative corrections uncertainty is not purely theoretical, but influenced by the experimental energy resolution. The latter can be tested by comparing the radiative tail in data and MC, or by varying the cut-off and inspecting the stability of the extracted ratio. 

In any case, the very definition of hard TPE is dependent on the employed radiative correction formalism. It is therefore of utmost importance to provide final data in a way that allows a post-publication update of the prescription, for example by tracking and publishing important kinematic parameters per bin. We further plan to publish the results relative to the most accepted theoretical treatments at the time.

\subsection{Charge-dependent inelastic background}

Invariably, some fraction of inelastic background will pass all applied cuts. In first order, this will dilute the ratio, a negligible effect. It is possible that the background is highly charge-dependent. This can be tested, and then corrected for, by investigating the extracted ratio for side-bands of the co-planarity plots. Such tests for OLYMPUS at 2 GeV beam gave negligible corrections. Additionally, such an effect can be modelled in MC with suitable generators. 

A second source of background is from scattering off the target walls. While elastic scattering can be separated easily via the measured momentum, quasielastic and radiative events will be part of the signal sample. Empty target measurements will be taken to calibrate MC simulations for a correction. It should be noted that the dominant quasielastic scattering of the protons in the wall is similar enough in kinematics and process to elastic lepton-proton scattering that, even if not corrected for, should lead to essentially the same TPE contribution and thus, $R_{2\gamma}$.

\subsection{Relative and absolute luminosity}

The requirements for the control of the relative luminosity are given by the effect size. To establish TPE as the major source of the form factor discrepancy, one has to measure at small $\epsilon$, where we expect the magnitude of the effect to reach tens of percent, and a relative luminosity uncertainty of even 1\% would be adequate. More demanding is the study of the ``dip" below 1 at high $\epsilon$, which requires control on the per-mille level.  

The existing Hall B set up is good to 2\% absolute determinations of the electron beam current \cite{CLAS:2015uuo}. The overall low beam current limits target effects like boiling, and we expect that target-related effects cancel in the ratio. We believe that the charge dependency of the current measurement will be minor. There may be slight differences due to the beam profile, or beam steering. a final analysis of the expected uncertainty will have to wait until the upgraded beam current monitoring system is implemented. 

As a fallback for the case that the luminiosity system fails to provide a reliable luminosity estimate, we can make use of the same trick employed by VEPP-3: our measurements reach to $\epsilon$ close to 1, where any TPE effect must vanish. This is easily good enough to determine the relative luminosity for the extraction at low $\epsilon$ at all three beam energies. For the lowest beam energy, the predicted minimum  of the dip is at sufficiently smaller $\epsilon$ to test this as well, while for the higher two beam energies, the measured $\epsilon$ range starts too close to the minimum so that the dip can be only tested via the curvature of the extracted functional form.

As discussed before, a determination of the absolute luminosity is not required for the core goal of this measurement. Nevertheless, an absolute estimation to a few \% is adequate to produce an impactful charge-averaged result. This is likely within in the capabilities of the BCM system.

\section{Elastic \texorpdfstring{$e^{-}p$}{e-p} coincidence with CLAS12}

\begin{figure}[h!]
    \centering
    \includegraphics[width = 0.5\textwidth]{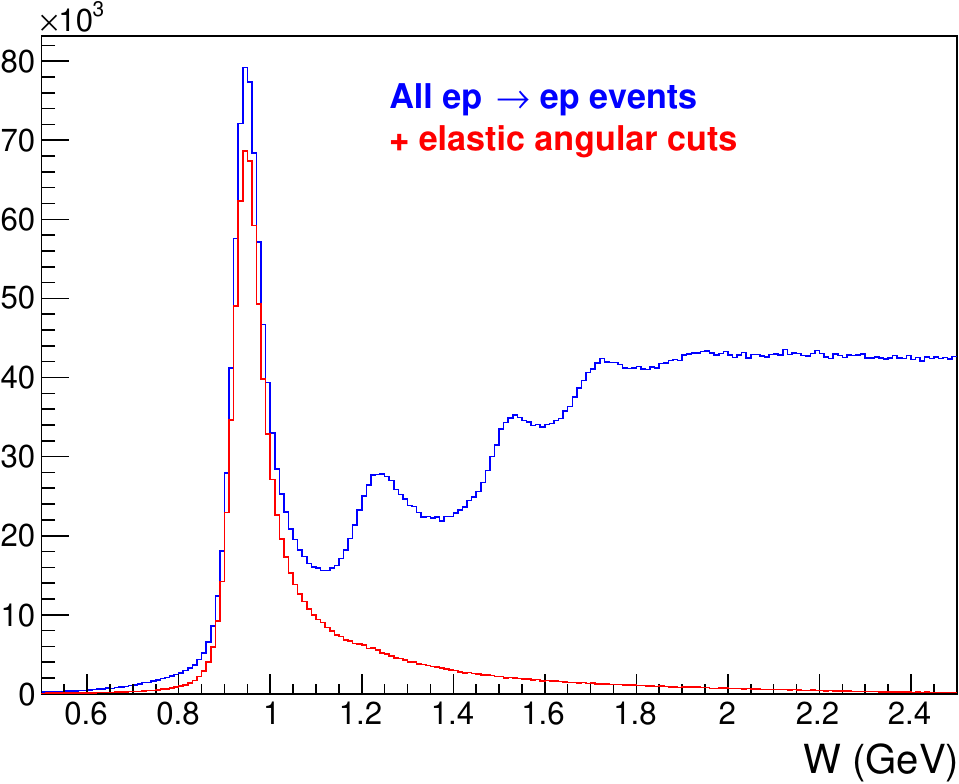}
    \caption{Invariant mass spectrum from 6 GeV $ep$ data collected by RG-M, for all $ep$ final states events (blue) and with elastic events isolated with cuts on the correlated electron and proton angle (red)}
    \label{fig:rgm_w}
\end{figure}

Coincident detection of scattered electrons and recoil protons in CLAS12 was assessed using data collected by the recent Run Group M (RG-M) experiment in Hall B.  The examined data was collected with a 6 GeV electron beam incident on a LH$_2$ target.  The invariant mass spectrum of the examined data is shown in Fig.~\ref{fig:rgm_w}, which shows both the full spectrum for all $ep\to ep$ events, along with the spectrum for elastic events identified with cuts on the correlated electron and proton angles (discussed below).

In elastic scattering, the electron and proton have back-to-back azimuthal angles ($\phi$), and their polar angles ($\theta$) are related by

\begin{equation}
    \theta_e(\theta_p) = 2 \arctan\left(\frac{1}{(1 + E_b/M_p)\tan\theta_p} \right).
    \label{eq:theta}
\end{equation}
\begin{figure}
    \centering
    \includegraphics[width = 0.54\textwidth]{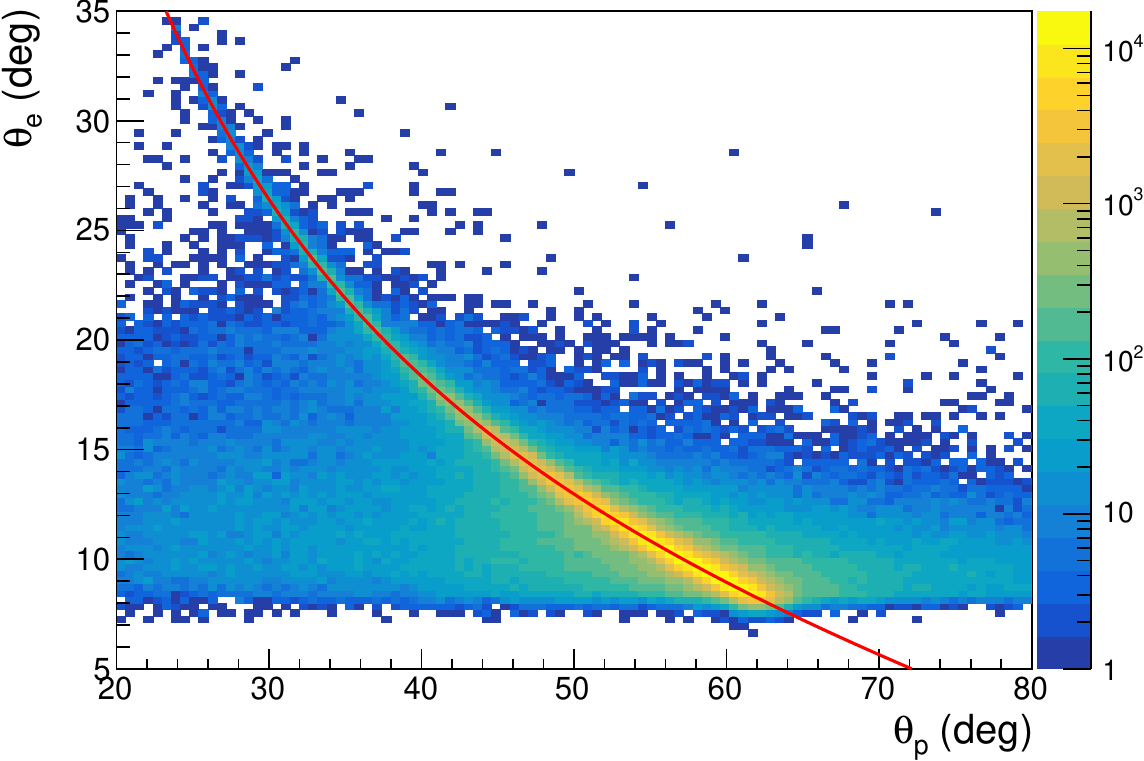}
    \caption{Polar angles of the electron and proton in elastic $ep$ events from RG-M data}
    \label{fig:rgm_ep_theta}
\end{figure}

The polar angles of the electron and proton for elastic events ($0.7 < W < 1.1$ GeV) are shown in Fig.~\ref{fig:rgm_ep_theta}, and are in good agreement with Eq.~\ref{eq:theta} (plotted in red).  

\newpage

\begin{figure}
    \centering
    \includegraphics[width = 0.45\textwidth]{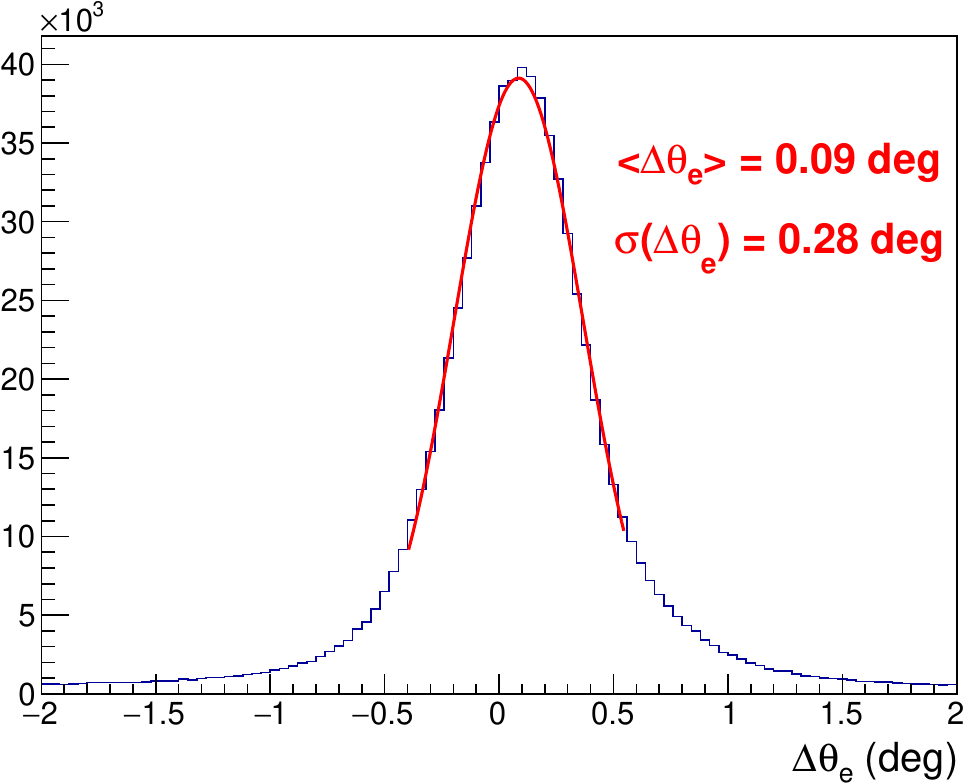}
    \hspace{25pt}
    \includegraphics[width = 0.45\textwidth]{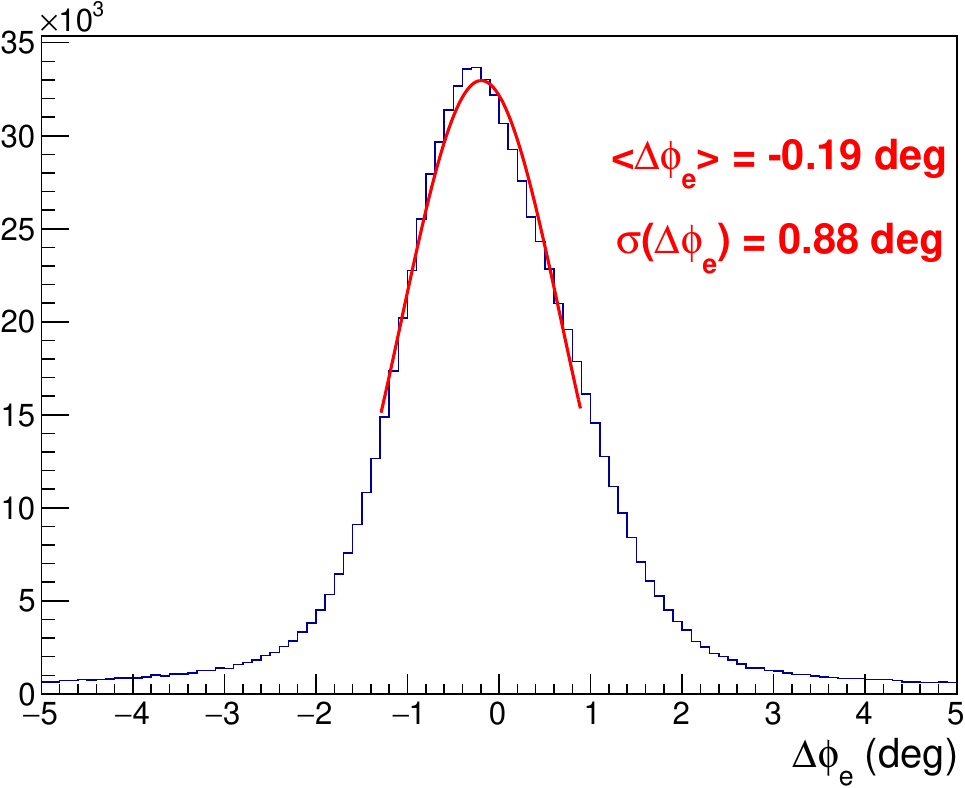}
    \caption{Resolution of the electron polar angle (left) and azimuthal angle (right), as defined by Eqs.~\ref{eq:dtheta} and \ref{eq:dphi}}
    \label{fig:dangle}
\end{figure}

Elastic events can therefore be identified by placing cuts on the correlated angles of the electron and proton.  The resolutions with which the electron angles can be determined by the measured proton are quantified by:
\begin{equation}
    \Delta\theta_e = \theta_e^{meas} - \theta_e(\theta_p)
    \label{eq:dtheta}
\end{equation}
\vspace{1pt}
\begin{equation}
    \Delta\phi_e = |\phi_e^{meas} - \phi_p^{meas}| - 180^\circ
    \label{eq:dphi}
\end{equation}
where $\theta^{meas}$ and $\phi^{meas}$ are the measured angles, and $\theta_e(\theta_p)$ is given by Eq.~\ref{eq:theta}.  These quantities, shown in Fig.~\ref{fig:dangle}, give 1$\sigma$ widths of $0.28^\circ$ and $0.88^\circ$ in $\theta$ and $\phi$, respectively.  To demonstrate the ability of angular cuts to identify elastic events, approximately $3\sigma$ cuts of $|\Delta\theta_e| < 1^\circ$ and $|\Delta\phi_e| < 3^\circ$ are applied.  The $W$ distribution for events satisfying these cuts is shown in red in Fig.~\ref{fig:rgm_w}, showing a clean isolation of elastic events. 

It should be noted that the CLAS12 trigger limits the RGM data to events with an electron going to the forward detector and a proton going to the central detector. We expect that the alternate topology (central lepton, forward proton) should have similar angle correlation resolutions.

\section{Run Plan and Projected Sensitivity}

We propose a 55 PAC day experiment in order to measure $R_{2\gamma}$ over a wide kinematic range with sub-percent statistical precision. We divide this time between three beam energies: standard first pass ($\approx$2.2 GeV), second pass ($\approx$4.4 GeV), and third pass ($\approx$6.6 GeV) energies, spending most of the time measuring at third pass, where the elastic scattering rates are lowest. We reiterate that the exact energy values are not critical; we will assume 2.2, 4.4, and 6.6 GeV in our estimates of statistical precision. We do not see a justification for measuring at fourth-pass. Such a measurement would require significantly more time to achieve reasonable precision. Third-pass running already reaches $Q^2=10$~GeV$^2$, a point beyond which we do not have a clear picture of the size of the form factor discrepancy. 

\subsection{Measurement Plan}

\begin{table}[htpb]
    \centering
    \begin{tabular}{c c c c c c}
    \hline
    \hline
    Energy & Species & Solenoid & Toroid & Target & PAC Days \\
    \hline
    2.2 & $e^-$ & $\leftarrow$ & $e^-$ in & empty & 0.083 \\
    2.2 & $e^-$ & $\leftarrow$ & $e^-$ in & full & 0.167 \\
    \hline
    \multicolumn{5}{c}{Polarity change} & 0.167 \\
        \hline
    2.2 & $e^-$ & $\leftarrow$ & $e^-$ out & full & 0.167 \\
        \hline
    \multicolumn{5}{c}{Polarity change} & 0.167 \\
        \hline
    2.2 & $e^-$ & $\rightarrow$ & $e^-$ out & full & 0.167 \\
        \hline
    \multicolumn{5}{c}{Polarity change} & 0.167 \\
        \hline
    2.2 & $e^-$ & $\rightarrow$ & $e^-$ in & full & 0.167 \\
        \hline
    \multicolumn{5}{c}{Species change} & 0.333 \\
        \hline
    2.2 & $e^+$ & $\rightarrow$ & $e^-$ in & full & 0.167 \\
        \hline
    \multicolumn{5}{c}{Polarity change} & 0.167 \\
        \hline
     2.2 & $e^+$ & $\rightarrow$ & $e^-$ out & full & 0.167 \\
         \hline
    \multicolumn{5}{c}{Polarity change} & 0.167 \\
        \hline
     2.2 & $e^+$ & $\leftarrow$ & $e^-$ out & full & 0.167 \\
         \hline
    \multicolumn{5}{c}{Polarity change} & 0.167 \\
        \hline
    2.2 & $e^+$ & $\leftarrow$ & $e^-$ in & empty & 0.083 \\
    2.2 & $e^+$ & $\leftarrow$ & $e^-$ in & full & 0.167 \\
        \hline
    \multicolumn{5}{c}{Pass Change} & 0.333 \\
        \hline
        \hline
    \multicolumn{5}{l}{Total $e^+$ Production Running } & 0.667 \\
    \multicolumn{5}{l}{Total $e^-$ Production Running } & 0.667 \\
    \multicolumn{5}{l}{Configuration Changes} &  1.667 \\
    \multicolumn{5}{l}{Calibrations} &  0.167 \\
    \hline
    \hline
    \multicolumn{5}{l}{Total} & 3.167 \\
    \hline
    \end{tabular}
    \caption{Proposed measurement plan for 2.2 GeV running}
    \label{tab:plan2}
\end{table}

\begin{table}[htpb]
    \centering
    \begin{tabular}{c c c c c c}
    \hline
    \hline
    Energy & Species & Solenoid & Toroid & Target & PAC Days \\
    \hline
    4.4 & $e^+$ & $\rightarrow$ & $e^-$ in & empty & 0.083 \\
    4.4 & $e^+$ & $\rightarrow$ & $e^-$ in & full & 0.500 \\
    \hline
    \multicolumn{5}{c}{Polarity change} & 0.167 \\
        \hline
    4.4 & $e^+$ & $\rightarrow$ & $e^-$ out & full & 0.500 \\
        \hline
    \multicolumn{5}{c}{Polarity change} & 0.167 \\
        \hline
    4.4 & $e^-$ & $\leftarrow$ & $e^-$ out & full & 0.500 \\
        \hline
    \multicolumn{5}{c}{Polarity change} & 0.167 \\
        \hline
    4.4 & $e^-$ & $\leftarrow$ & $e^-$ in & full & 0.500 \\
        \hline
    \multicolumn{5}{c}{Species change} & 0.333 \\
        \hline
    4.4 & $e^+$ & $\leftarrow$ & $e^-$ in & full & 0.500 \\
        \hline
    \multicolumn{5}{c}{Polarity change} & 0.167 \\
        \hline
     4.4 & $e^+$ & $\leftarrow$ & $e^-$ out & full & 0.500 \\
         \hline
    \multicolumn{5}{c}{Polarity change} & 0.167 \\
        \hline
     4.4 & $e^+$ & $\rightarrow$ & $e^-$ out & full & 0.500 \\
         \hline
    \multicolumn{5}{c}{Polarity change} & 0.167 \\
        \hline
    4.4 & $e^+$ & $\rightarrow$ & $e^-$ in & empty & 0.083 \\
    4.4 & $e^+$ & $\rightarrow$ & $e^-$ in & full & 0.500 \\
        \hline
    \multicolumn{5}{c}{Pass Change} & 0.333 \\
        \hline
        \hline
    \multicolumn{5}{l}{Total $e^+$ Production Running } & 2 \\
    \multicolumn{5}{l}{Total $e^-$ Production Running } & 2 \\
    \multicolumn{5}{l}{Configuration Changes} &  1.667 \\
    \multicolumn{5}{l}{Calibrations} &  0.167 \\
    \hline
    \hline
    \multicolumn{5}{l}{Total} & 5.833 \\
    \hline
    \end{tabular}
    \caption{Proposed measurement plan for 4.4 GeV running}
    \label{tab:plan4}
\end{table}

\begin{table}[htpb]
    \centering
    \begin{tabular}{c c c c c c}
    \hline
    \hline
    Energy & Species & Solenoid & Toroid & Target & PAC Days \\
    \hline
    6.6 & $e^-$ & $\leftarrow$ & $e^-$ out & empty & 0.083 \\
    6.6 & $e^-$ & $\leftarrow$ & $e^-$ out & full & 5.500 \\
    \hline
    \multicolumn{5}{c}{Species change} & 0.333 \\
        \hline
    6.6 & $e^+$ & $\leftarrow$ & $e^-$ out & full & 5.500 \\
        \hline
    \multicolumn{5}{c}{Polarity change} & 0.167 \\
        \hline
    6.6 & $e^+$ & $\leftarrow$ & $e^-$ in & full & 5.500 \\
        \hline
    \multicolumn{5}{c}{Species change} & 0.333 \\
        \hline
    6.6 & $e^-$ & $\leftarrow$ & $e^-$ in & empty & 0.083 \\
    6.6 & $e^-$ & $\leftarrow$ & $e^-$ in & full & 5.500 \\
        \hline
    \multicolumn{5}{c}{Polarity change} & 0.167 \\
        \hline
    6.6 & $e^-$ & $\rightarrow$ & $e^-$ in & full & 5.500 \\
        \hline
    \multicolumn{5}{c}{Species change} & 0.333 \\
        \hline
    6.6 & $e^+$ & $\rightarrow$ & $e^-$ in & full & 5.500 \\
         \hline
    \multicolumn{5}{c}{Polarity change} & 0.167 \\
        \hline
    6.6 & $e^+$ & $\rightarrow$ & $e^-$ out & full & 5.500 \\
         \hline
    \multicolumn{5}{c}{Species change} & 0.333 \\
        \hline
    6.6 & $e^-$ & $\rightarrow$ & $e^-$ out & full & 5.500 \\
        \hline
        \hline
    \multicolumn{5}{l}{Total $e^+$ Production Running } & 22 \\
    \multicolumn{5}{l}{Total $e^-$ Production Running } & 22 \\
    \multicolumn{5}{l}{Configuration Changes} &  1.833 \\
    \multicolumn{5}{l}{Calibrations} &  0.167 \\
    \hline
    \hline
    \multicolumn{5}{l}{Total} & 46 \\
    \hline
    \end{tabular}
    \caption{Proposed measurement plan for 6.6 GeV running}
    \label{tab:plan6}
\end{table}

Tables~\ref{tab:plan2}, \ref{tab:plan4}, and \ref{tab:plan6} show our proposed measurement plan for the 2.2~GeV, 4.4~GeV, and 6.6~GeV running periods. The total time allocation is summarized in Table~\ref{tab:plan_summary}. 
The polarity of the solenoid magnet is indicated with a `$\rightarrow$' to indicate that the $\vec{B}$-field is parallel to the beam, and with a `$\leftarrow$' to indicate anti-parallel. The toroid polarity is labelled by whether electrons are bent inwards toward the beamline, or outward, away from the beamline. 

\begin{table}[htpb]
    \centering
    \begin{tabular}{l|l}
\hline\hline
     Setting  &  PAC Days \\
     \hline
      2.2 GeV production & 1.333 \\
      4.4 GeV production & 4 \\
      6.6 GeV production & 44 \\
      Calibrations & 0.5 \\
      Configuration Changes & 5.167 \\
      \hline
      Totals & 55\\
      \hline
    \end{tabular}
    \caption{A summary of the time allocation requested in the run plan}
    \label{tab:plan_summary}
\end{table}

We assume that pass changes and beam species changes can be accomplished in 0.333 PAC days ($\approx 2$ shifts), while a change in polarity in one of the two CLAS12 magnets can be accomplished in 0.167 PAC days ($\approx 1$ shift). 

For the 2.2 GeV and 4.4 GeV run periods, the total production time per target is so short, there is little advantage conferred by interleaving electron and positron data taking. We assume that the runs are short compared to the time-scale over which the detector performance is relatively stable. For the longer 6.6 GeV running, we interleave the electron and positron runs. 

We take a small amount of empty target data for both beam species but for one magnet configuration. For this experiment, there is no need to perform a target-wall background subtraction. Instead this data can be used to normalize a quasi-elastic Monte Carlo simulation, with which we can perform background studies. 

\begin{figure}[bht]
    \centering
    \includegraphics{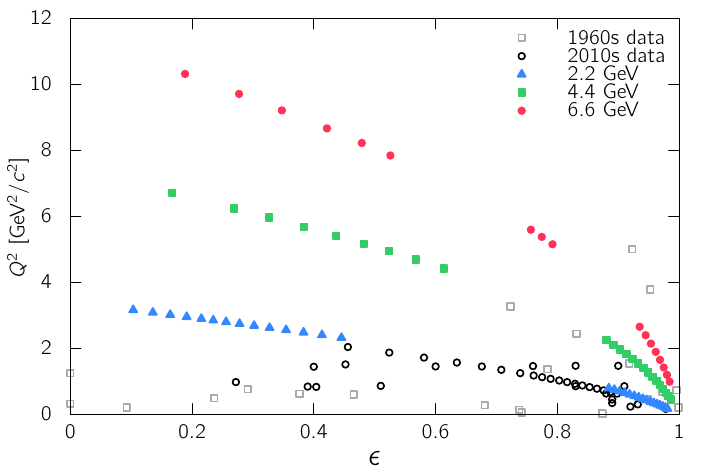}
    \caption{Comparison of kinematics of the data points in the proposed measurement and previous $R_{2\gamma}$ data}
    \label{fig:proj_kin}
\end{figure}

\subsection{Rate Estimates}

We have performed an estimate of the expected statistical precision that could be obtained from this run plan. For these estimates, we have assumed a luminosity of $10^{35}$ cm$^{-2}$s$^{-1}$, a fully efficient trigger, and a fully efficient detector within the fiducial acceptance. 

For estimation purposes we have made a simplified model of the CLAS12 fiducial acceptance. 
For the central detector, we have assumed full $2\pi$ azimuthal coverage between polar angles of $40^\circ$ and $120^\circ$, subject to a momentum threshold of $p_T > 350$~MeV$/c$. 
For the forward detector, the forward acceptance is limited by in-bending particles, while the backward acceptance is limited by out-bending particles. Therefore, the angular ranges of the fiducial region are momentum dependent. 
In order to perform an analysis with all four magnetic field configurations, all particles, for a given kinematics, must be accepted under both in-bending and out-bending toroid polarities. 
We approximate the forward in-bending acceptance as $\theta > 7^\circ + 8^\circ \cdot \text{GeV}/c / p$, where $p$ is the in-bending particle momentum. 
Similarly, we approximate the backward out-bending acceptance as $\theta < 31^\circ - 8^\circ\cdot \text{GeV}/c / p$. 
We approximate that the forward detector azimuthal coverage changes from linearly from 0 at the minimum polar angle for a given momentum to a $\pm 25^\circ$ per sector over the next $5^\circ$ of polar angle. 
This highly approximate model can no doubt be improved by detailed GEMC Monte Carlo simulations in the future.

For our estimates, we only consider kinematics in which CLAS12 has acceptance for all particles in all four field settings, as this is the most systematically favorable measurement that can be made. However, with extra effort, one could analyze events in kinematics with full acceptance in only two out of the four magnet settings, or even with only one of the two out-going particles detected. Such analyses may extend the kinematic reach beyond what we present here.

For a cross section, we assume a Born-level Rosenbluth cross section, with the form factors given by the global fit to unpolarized data given in Ref.~\cite{A1:2013fsc}. 

From our rate estimates, we developed a binning scheme with the goal of having approximately 1\% statistical precision or better per bin. The kinematics of these bins are shown in comparison to the kinematics of previous $R_{2\gamma}$ data points in Fig.~\ref{fig:proj_kin}. The previous data are the same as those shown in Fig.~\ref{fig:prev_data_kin}. The kinematic reach toward both low-$\epsilon$ and high-$Q^2$ far exceeds previous measurements.

\begin{figure}
    \centering
    \includegraphics[width=0.45\textwidth]{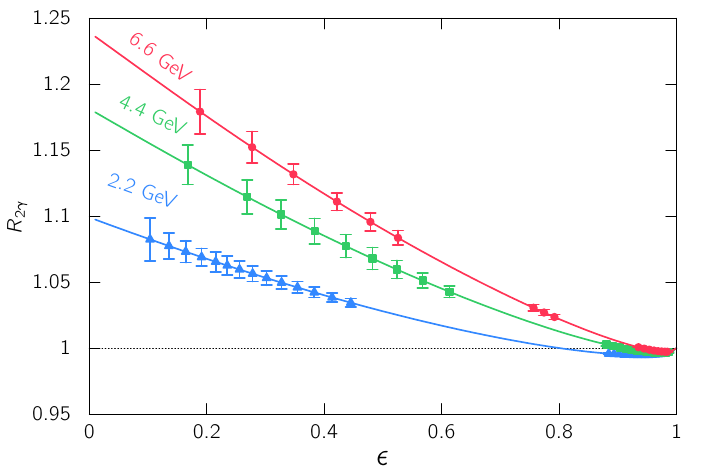}\hspace{0.05\textwidth}
        \includegraphics[width=0.45\textwidth]{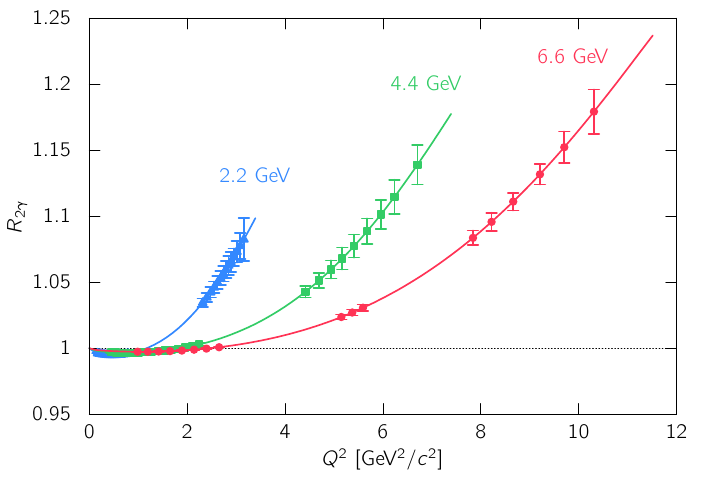}   
    \caption{Estimated statistical uncertainty on TPE as a function of $\epsilon$ (left) and $Q^2$ (right).  The data points are placed on the curves of the phenomenological TPE estimate of Ref.~\cite{A1:2013fsc}.}
    \label{fig:proj_unc}
\end{figure}

Our projected statistical uncertainties are shown in Fig.~\ref{fig:proj_unc} as a function of $\epsilon$ (left) and $Q^2$ (right). The data points are placed on the curves of the phenomenological TPE estimate of Ref.~\cite{A1:2013fsc}. These curves indicate the approximate effect size in order to explain the measured proton form factor discrepancy. The precision will be sufficient to make a definitive test of the TPE hypothesis, and place strong constraints on both the size and kinematic dependencies of $R_{2\gamma}$.

In both Figs.~\ref{fig:proj_kin}, and \ref{fig:proj_unc}, there are several gaps in the data, which occur at the interface between the forward and central detectors of CLAS12. The points at the highest $\epsilon$ (smallest $Q^2$) correspond to event topologies where the lepton scatters into the forward detector and the proton scatters into the central detector. The lower $\varepsilon$ (higher $Q^2$) points are event topologies where the lepton scatters into the central detector and the proton scatters into the forward detector. It is worth noting that at the 6.6 GeV beam energy, there are three data points where both the lepton and proton scatter into the forward detector. If either the lepton or proton does not scatter into the fiducial acceptance of either detector in our simplified acceptance model for even one field configuration, then this point is excluded. This occurs prominently in the central region of the $\epsilon,Q^2$ plane. We note that detailed Monte Carlo studies will give a better indication of the exact acceptance edges of the detectors. For this proposal we have made conservative assumptions. 

A future positron-capabilities at CEBAF provide a clear opportunity to make a high-precision determination of the hard TPE effect in elastic electron-proton scattering over a wide phase-space. For this effort, we request 55 PAC days.

\section{Relationship to Other Approved Experiments}

There are no other approved experiments that will directly measure hard TPE using positrons at this time.

There are two conditionally approved experiments that use the planned positron beam capabilities. E12-20-009 (CLAS12)
and E12-20-012 (Hall C) will use positrons to measure deeply virtual Compton scattering. 

There are several approved experiments that relate to two-photon exchange in some way. 
\begin{itemize}
    \item E12-22-004 (SOLID) will look at the beam-normal single spin asymmetry in DIS, which probes the imaginary part of the TPE amplitude in DIS.
    It should be noted that the proton form factor discrepancy is caused by the real part of the TPE amplitude, and is insensitive to the imaginary part.
    \item The nucleon form factor experiments of the Super BigBite program in Hall A, will help clarify the extent of the form factor discrepancy at even higher values of $Q^2$. Specifically
    \begin{itemize}
        \item E12-09-019, in combination with E12-20-010, will determine the neutron form factor ratio $\mu_n G_E^n/G_M^n$ through unpolarized quasi-elastic scattering from deuterium. This can be compared with the results from polarization observables to establish if there is a similar form factor discrepancy for neutrons. 
        \item E12-09-016 and E12-17-004 will measure the neutron form factor ratio $\mu_n G_E^n/G_M^n$ using polarization observables.
        \item E12-07-109 will be the highest-$Q^2$ measurement of $\mu_p G_E^p/G_M^p$, and will provide a first glimpse of the how the proton form factor discrepancy evolves beyond $Q^2=10$~GeV$^2/c^2$.
    \end{itemize}
\end{itemize}

\bibliographystyle{JHEP}
\bibliography{references.bib}

\end{document}